# On the vibrational properties of transition metal doped ZnO: surface, defect, and bandgap engineering

Viviane M. A. Lage[a,b], Carlos Rodríguez-Fernández[c,d], Felipe S. Vieira[b],

Rafael T. da Silva[b], Maria Inês B. Bernardi[e], Maurício M de Lima Jr.[f],

Andrés Cantarero[c] and Hugo B. de Carvalho[b*]

[a] *Universidade Federal de Ouro Preto – UFOP, 35400-000 Ouro Preto, MG, Brazil*

[b] *Instituto de Ciências Exatas, Universidade Federal de Alfenas – UNIFAL, 37130-000 Alfenas, Brazil.*

[c] *Institute of Molecular Science, University of Valencia, 22085, E-46071 Valencia, Spain.*

[d] *Faculty of Engineering and Natural Sciences, Tampere University, 33720 Tampere, Finland*

[e] *Instituto de Física de São Carlos, Universidade de São Paulo – USP, 13560-970 São Carlos, Brazil*

[f] *Materials Science Institute (ICMUV), University of Valencia, 22085, E-46071 Valencia, Spain.*

*Corresponding Author*:
* hugo.carvalho@unifal-mg.edu.br

**ABSTRACT:** We present a comprehensive study on the structure and optical properties of Mn- and Co-doped ZnO samples prepared via solid-state reaction method with different dopant concentrations and atmospheres. The samples were structural and chemically characterized via X-ray diffraction, scanning electron microscopy, energy-dispersive X-ray spectroscopy, and X-ray excited photoelectron spectroscopy. The optical characterization was performed via Raman, photoluminescence, and diffuse photoreflectance spectroscopies. Emphasis was done on the studies of their vibrational properties. The structural data confirm the incorporation of Mn and Co ions into the wurtzite ZnO lattice. It is demonstrated that the usual observed additional bands in the Raman spectrum of transitional metal (TM) doped ZnO are related to structural damage, deriving from the doping process, and surface effects. The promoted surface optical phonons (SOP) are of Fröhlich character and, together with the longitudinal optical (LO) polar phonons, are directly dependent on the ZnO electronic structure. The enhancement of SOP and LO modes with TM-doping is explained in terms of nonhomogeneous doping, with the dopants concentrating mainly on the surface of grains, and a resonance effect due to the decrease of the ZnO bandgap promoted by the introduction of the 3$d$ TM levels within the ZnO bandgap. We also discuss the origin of the controversial vibrational mode commonly observed in the Mn-doped ZnO system. It is stated that the observation of the analyzed vibrational properties is a signature of substitutional doping of the ZnO structure with tuning of ZnO optical absorption into the visible range of the electromagnetic spectrum.

**KEYWORDS:** TM-doped ZnO, vibrational properties, surface, defect, and bandgap engineering.

## 1 INTRODUCTION

Oxide-based semiconductors correspond to a broad class of tunable materials with promising applications in different traditional systems and emerging technologies [1, 2]. Their functionalization is usual accomplished through defect engineering: by introducing intrinsic defects into the structural lattice in a well-controlled manner or via doping [3, 4]. Small concentrations of dopant elements can significantly change the electronic properties of oxide semiconductors. Dopants can also induce local and gradient strains, built-in electrostatic fields, novel magnetic phases; dopants may occupy different regular lattice or interstitial sites with drastically different consequences for the physical properties of the host material [5, 6]. Additionally, due to the huge correlation among, charge, spin, and structure characteristics presented in the oxide semiconductors, several new exciting defect/doping effects may be present in such materials, like insulator-metal transitions [7], interfacial/surface [8-10] and bulk [11-13] room temperature ferromagnetism, and superconductivity [14, 15]. A careful characterization of the dopant atoms is crucial to master the complexity and understand the emergent composition-driven phenomena and opportunities in oxide materials. For this purpose, different experimental techniques have been employed providing complementary information on the spatial doping distribution [1]. In such a context, Raman spectroscopy is a valuable tool; it is a very sensitive spectroscopy technique that returns important information about the nature of the solid on a scale of a few nanometers [3]. Therefore, Raman spectroscopy can be used to study the microscopic nature of structural and/or morphological disorders, which are strongly correlated with optical phonons [4]. Raman spectroscopy is also a powerful tool to identify various phases in polycrystalline inorganic materials and in studying underlying atomic and electronic structure changes caused by doping [16, 17].

Among the oxide semiconductors, zinc oxide (ZnO) deserves the distinction. ZnO is a multifunctional material presenting distinctive properties, such as high electron mobility, high thermal conductivity, wide direct bandgap (~3.3 eV), and large exciton binding energy (~60 meV) [18]. The unique and tunable ZnO properties show excellent chemical stability as well as thermally stable $n$-type semiconducting with wide applications such as luminescent material, in supercapacitors, batteries, solar cells, photocatalysis, biosensors, biomedical and biological applications in the form of pellets, thin film, and nanoparticles [19]. ZnO crystalizes in the wurtzite structure, with the $Zn^{2+}$ atoms occupying tetrahedral sites. In contrast, the octahedral interstitial sites are empty, letting many available lattice sites to house defects and dopants. In particular, the ZnO doping with $3d$ transition metals (TM) is quite interesting. The $3d$ TM elements can assume single and multiple oxidation states, whose effects on the physical properties are significant [20]. TM-doped ZnO has been explored in different technological areas, especially in developing materials for application in spintronic devices, dilute magnetic semiconductors (DMSs), and photocatalytic systems (heterogeneous photocatalysis). In spintronics, the development of DMSs is one of the most controversial research topics in materials science. Regarding the nature of its ferromagnetic properties, excluding the trivial extrinsic origins due to ferromagnetic secondary phases, the main theoretical models that are currently available are all linked to structural defects [11-13, 17, 21, 22]. Therefore, the study and determination of the phase diagram between ZnO and TM oxides are essential to assure the synthesis of true TM-doped ZnO ($Zn_{1-x}TM_xO$ alloys) with no secondary/spurious undesired phases. Under ambiental concerns, ZnO is an abundant, nontoxic resource with superior environmental affinity, making it an exciting candidate for heterogeneous photocatalysis [23-27]. The main challenge in developing a ZnO-based photocatalyst is to tune its optical absorption from UV to the visible part of the solar spectrum. Here, the strategy is to add specific energy levels within the ZnO bandgap via doping and/or by introducing proper structural point defects in the system [28]. ZnO has also been emerging as one of the potential materials in solar cell applications (photovoltaic systems) owing to its high conductivity, high electron affinity, excellent electron mobility, stability against photo-corrosion, and availability at low-cost [29]. Doping ZnO with different elements has proved to be an efficient way to improve its exciton lifetime [30], electronic properties [31, 32], and optical absorption by decreasing its energy bandgap [33-35].

Considering the vibrational properties of TM-doped ZnO, the emergence of additional modes and broad bands in the ZnO Raman spectrum between 450 and 650 $cm^{-1}$ is often observed under doping. The appearance of these additional modes and bands and their dependence on the dopant concentration is observed for different doping elements, for instance: H [36], N [36, 37], P [38], Mn [39], Co [21], Ni [40], Cu [41], Ga [37], Ag [42], and Sb [43]. There are also reports showing the activation of these modes in undoped ZnO via mechanical milling [44, 45]. In ion-implanted ZnO samples where, besides the dopant incorporation, the bombardment process leads to significant densities of structural defects, after performing an annealing process removing the structural defects, the broadband completely disappears [37, 38]. These results lead us to infer that the observation of these modes is related to structural defects introduced in the ZnO lattice due to the dopant incorporation or extrinsic structural damage. In fact, the indexation of these additional vibrational modes and bands in this region of the ZnO Raman spectrum is a highly controversial issue.

Facing these contexts, we present a comprehensive study on the ZnO vibrational properties of Mn- and Co-doped ZnO. The present report aims to give a contribution to the understanding of the origin of the observed additional modes and bands in the ZnO Raman spectrum under doping and/or subject to structural damage. Our results demonstrate that the additional modes and bands are closely related to surface effects and the electronic structure of the system, giving valuable information about the dopant incorporation and the structural defect concentrations. This knowledge opens up further opportunities in the ZnO surface, defect, and bandgap engineering for ZnO functionalization and its application in many important technological devices.

## 2 EXPERIMENTAL METHODS

Polycrystalline $Zn_{1-x}Mn_xO$ samples with Mn nominal concentrations of $x_N = 0.002, 0.005, 0.010, 0.015$, and $0.020$ ($0.2, 0.5, 1, 1.5$, and $2$ at.%), and $Zn_{1-x}Co_xO$ samples with Co nominal concentrations of $x_N = 0.01, 0.03, 0.05, 0.07$, and $0.09$ ($1, 3, 5, 7$, and $9$ at.%) were prepared via standard solid-state reaction method. Stoichiometric amounts of ZnO (Sigma Aldrich 99.99% purity), metallic Mn (Sigma Aldrich 99.9% purity), and $Co_3O_4$ (Sigma Aldrich 99.7% purity) powders were mixed for 4 h at 200 rpm in a planetary ball mill. The resulting powder was cold compacted at a pressure of 200 MPa in the form of pellets (green pellets). The green pellets were finally sintered at 1200 °C for 4 hours. The $Zn_{1-x}Mn_xO$ set of samples was sintered in argon (Ar) atmosphere, flux of 1.5 l/min, while the $Zn_{1-x}Co_xO$ set was sintered in oxygen ($O_2$), 1.5 l/min. The specific sintering atmospheres for Mn (Ar) and Co ($O_2$) were chosen to maximize the solubility limit for each dopant in the wurtzite ZnO structure. These choices are based on our previous studies [16, 17, 21, 46]. For each set of samples, an undoped ZnO reference sample was prepared in the same condition as the doped ones.

An additional set of mechanically milled $Zn_{1-x}Co_xO$ ($x_N = 0.05$) samples was also prepared. The powder mixture (ZnO and $Co_3O_4$) was not pressed into pellets and was sintered under the same conditions as the previously presented bulk samples. After sintering, the powder was subjected to high-energy mechanical milling in a planetary tungsten carbide ball mill (RETSCH PM100) at a rotation speed of 500 rpm. A total of 20 g of sample was mechanically milled, keeping the ratio between the masses of the spheres and the powder at ~13:1. Isopropyl alcohol was also added to the powder to optimize grinding. Fractions of the powder were collected after 6, 12, 18, and 24 h of milling.

X-ray diffraction (XRD), scanning electron microscopy (SEM), energy-dispersive X-ray spectroscopy (EDS), X-ray excited photoelectron spectroscopy (XPS), and Raman scattering spectroscopy were used to study the incorporation of Mn and Co and the resulting lattice disorder in the ZnO host structure. The XRD measurements were recorded in the range of $2\theta = 20°-80°$ with steps of $0.02°$ at 10 s/step by using Cu-K$\alpha$ radiation ($\lambda = 1.542$ Å) of a Rigaku Ultima IV diffractometer. The microstructure was determined using SEM with a LV JEOL JSM 6510 and a resolution of 3 nm at 30 kV. The elemental distribution and effective dopant concentrations ($x_E$) incorporated into the ZnO structure were estimated by EDS using an Oxford XMAX 50 detector. The analysis of the oxidation states of the elements present on the surface of the samples was performed by XPS using a Scientia Omicron ESCA+ spectrophotometer equipped with an Al-K$\alpha$ monochromatic X-ray source of 1486.7 eV with 280 W incident power and 50 eV constant pass power mode. The Raman spectra of the samples were performed at room temperature ($\sim 300$ K) and 10 K in backscattering configuration with a Jobin Yvon-Horiba T64000 triple spectrometer (gratings of 1800 grooves/mm). The excitation was performed using 351, 458, 488, 514, and 532 nm laser lines. The laser was focused onto the cryostat through the windows with lenses, giving a spot size of around 100 microns, which is relatively representative for statistical purposes. Diffuse photoreflectance measurements were realized in the range of 350–850 nm in a Lambda 1050

Perkin Elmer spectrophotometer. Changes in the density of defects in the mechanically milled samples were estimated by photoluminescence (PL) recorded at 5 K and excited using a He-Cd laser (325 nm).

## 3 RESULTS AND DISCUSSION

### 3.1 *Structural and chemical characterization*

Figure 1 presents the XRD patterns for the sets of samples, the vertical axes are in logarithm scale to highlight the potential presence of secondary phases, $ZnMn_2O_4$ for the $Zn_{1-x}Mn_xO$ set and $Co_3O_4$ for the $Zn_{1-x}Co_xO$ set of samples (symbols in Figure 1). Only peaks related to the wurtzite ZnO structure are observed (ICDD crystal chart no. 36-1451, space group $P6_3mc$), indicating, within the limit of detection of the technique, the absence of secondary phases and, consequently, the incorporation of the dopants into the ZnO lattice.

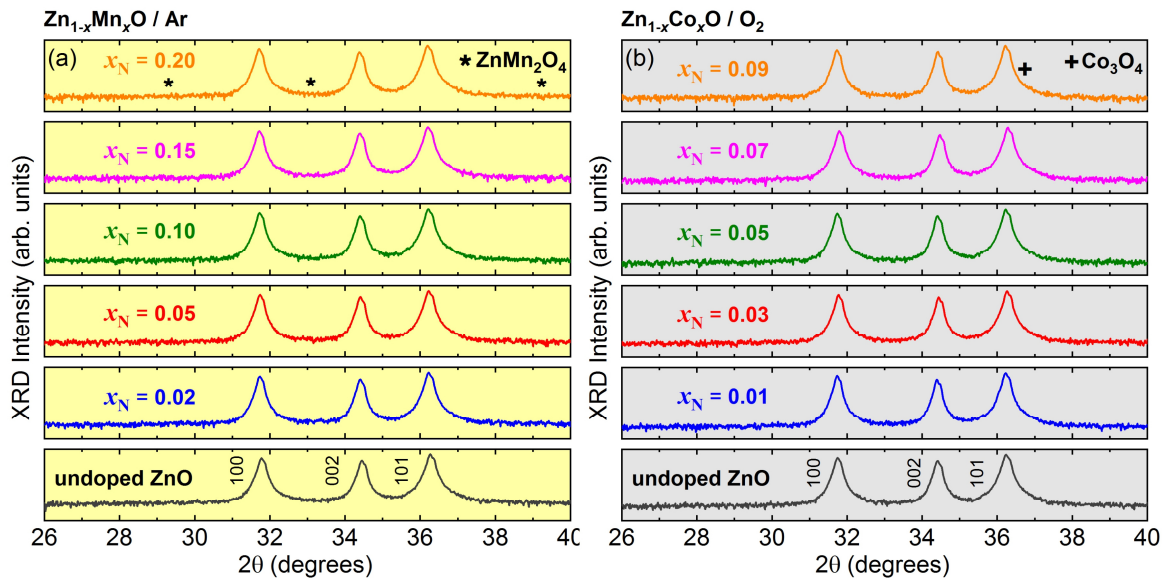

**Figure 1:** XRD pattern of the (a) $Zn_{1-x}Mn_xO$ ($x_N$ = 0.020, 0.015, 0.010, 0.005, 0.002, and 0.0 (undoped ZnO)) samples sintered in Ar; and of the (b) $Zn_{1-x}Co_xO$ ($x_N$ = 0.09, 0.07, 0.05, 0.03, 0.01, and 0.0 (undoped ZnO)) samples sintered in $O_2$. The symbols in the first panel correspond to the Bragg angle of the main usually observed secondary phases: (a) $ZnMnO_2$ for the Mn-doped ZnO system (PDF) and (b) $Co_3O_4$ for the Co-doped ZnO system. The vertical axes are in logarithmic scale to highlight the possible presence of secondary phases.

Figures 2(a) and 2(b) show a representative EDS spectrum and SEM micrograph with its correspondent elemental mapping obtained for the samples with higher dopant concentration, the $Zn_{1-x}Mn_xO$ ($x_N = 0.02$) and $Zn_{1-x}Co_xO$ ($x_N = 0.09$), respectively. The analyses were performed at multiple points and also over large areas. The images were obtained using a backscattered electron detector to highlight the presence of undesired secondary phases. It can be observed that the Mn and Co distributions are quite homogeneous, no dopant-rich regions were detected. These results indicate that the dopants in our samples are incorporated into the ZnO lattice. The measured average effective cobalt concentration ($x_E$) for the $Zn_{1-x}Mn_xO$ ($x_N = 0.02$) samples was $x_E = 0.021(3)$ and for the $Zn_{1-x}Co_xO$ ($x_N = 0.09$) sample was $x_E = 0.095(4)$. XPS measurements were also employed to evaluate the chemical composition and the elemental oxidation states on the surface of the samples. Figure S1 in the supplementary file presents the survey spectra and the high-resolution spectra around the $Zn\,2p$ and $O\,1s$ core level binding energies for the $Zn_{1-x}Mn_xO$ ($x_N = 0.02$) and $Zn_{1-x}Co_xO$ ($x_N = 0.09$) samples. No other impurities than carbon were detected (Figures S1(a) and S1(d)) [47]. The XPS at the $O\,1s$ binding energy reveals also the presence of chemisorbed oxygen species (Figures S1(c) and S1(f)) [47]. Figures 2(c) and 2(d) show the obtained XPS spectrum at the $Mn\,2p$ and $Co\,2p$ binding energies for the $Zn_{1-x}Mn_xO$ ($x_N = 0.02$) and $Zn_{1-x}Co_xO$ ($x_N = 0.09$) samples, respectively. In both cases four peaks are observed, corresponding to the $2p_{3/2}$ and $2p_{1/2}$ doublet and the shakeup resonance transitions (satellites) of these two peaks at higher binding energies. The energy for the $2p_{3/2}$ and $2p_{1/2}$ doublet obtained after fitting was 641.2 and 653.1 eV ($\Delta \approx 11.9$ eV) for the $Mn\,2p$, and 781 and 796.6 eV ($\Delta \approx 15.6$ eV) for the $Co\,2p$. The binding energies and energy separation ($\Delta$) of these peaks are characteristics of $Mn^{2+}$ and $Co^{2+}$ [22, 48, 49]. These oxidation states, together with the obtained XRD and EDS-mapping results, lead us to conclude that the dopants, Mn and Co, in the studied samples, are truly incorporated into the wurtzite ZnO structure. The Raman and diffuse reflectance results further presented also support this statement.

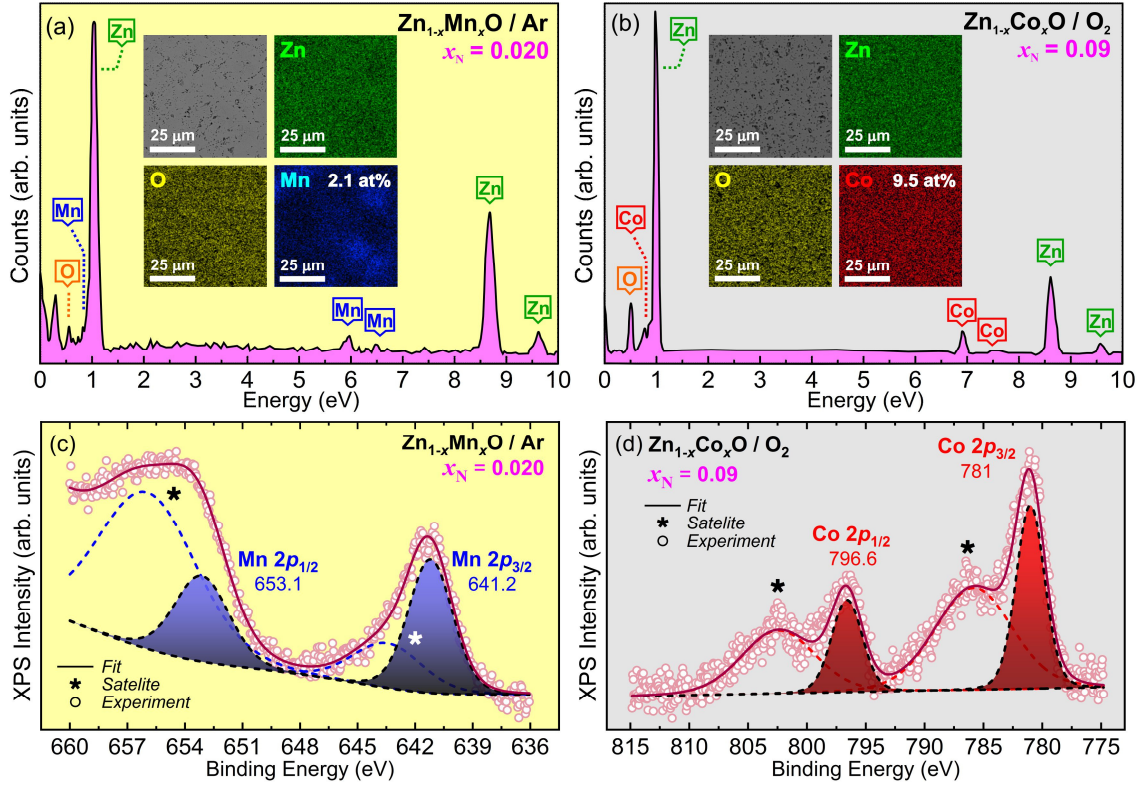

**Figure 2:** EDS spectra for the (a) $Zn_{1-x}Mn_xO$ ($x_N = 0.02$) and (b) $Zn_{1-x}Co_xO$ ($x_N = 0.09$) samples. The inset shows representative SEM micrographs and their EDS elemental mapping. High-resolution XPS spectra (c) at the $Mn\,2p$ core level energy for the $Zn_{1-x}Mn_xO$ ($x_N = 0.02$) sample and (d) at the $Co\,2p$ core level energy for the $Zn_{1-x}Co_xO$ ($x_N = 0.09$) sample.

### 3.2 *Raman as a function of the alloy content*

Figure 3 shows the Raman spectra obtained at 10 K with excitation of 458 nm for the $Zn_{1-x}Mn_xO$ (Figure 3(a)) and $Zn_{1-x}Co_xO$ (Figure 3(b)) sets of samples. The spectra are normalized by the integrated area of the $E_{2H}$ vibrational mode obtained after performing a Gaussian fit of the spectra. At the $\Gamma$-point of the Brillouin Zone (BZ) of the hexagonal wurtzite ZnO, the optical phonons are represented by $\Gamma = A_1 + 2B_1 + E_1 + 2E_2$. Here, $A_1$ and $B_1$ are one-dimensional modes, while $E_1$ and $E_2$ are two-dimensional modes. The two polar $A_1$ and $E_1$ modes and the two nonpolar $E_2$ modes are Raman active, while the two $B_1$ modes are inactive (silent modes). Thus, the $E_1$ and $A_1$ modes are split into transverse optical (TO) and longitudinal optical (LO) phonons [16, 17]. The $E_{2H}$ mode corresponds mainly to vibrations of the O sub-lattice, that in principle is not directly affected by the doping at the Zn sites. A Gaussian fit of the vibrational modes reveals to be more appropriate due to the alloying disorder. From here up to the end of the article, all the presented Raman spectra are normalized by the $E_{2H}$ integrated area. In Figures 3(a) and 3(b) the vertical scales are the same for comparison. Here, we observed for all samples the ZnO vibrational modes:

$2E_{2L}$ at the M-point of the BZ [16, 50] (~329 cm$^{-1}$), $A_1$ (TO) (~378 cm$^{-1}$), $E_1$ (TO) (~407 cm$^{-1}$), and $E_{2H}$ (~434 cm$^{-1}$) [51, 52]. Figure S1 presents the Raman spectra measured for the main Mn- and Co-related possible secondary phases in our samples. By comparison (Figures S1 and 2), any vibrational modes associated with secondary phases in the studied $Zn_{1-x}Mn_xO$ and $Zn_{1-x}Co_xO$ set of samples (Figure 2) are observed. This result supports the conclusion of a complete incorporation of the dopants, Mn and Co, into the wurtzite ZnO lattice.

For both sets of samples, we also observed the emergence of a broadband centered at ~475 cm$^{-1}$ (AB band) and the presence of additional vibrational modes around 500-550 cm$^{-1}$ (AMs), forming with the LO mode at ~575 cm$^{-1}$, another broadband ranging from 500-600 cm$^{-1}$. The LO mode corresponds to the $A_1$(LO) mode since the $E_1$(LO) can only be observed in 90° scattering configuration [51, 52]. In ZnO, the $A_1$(LO) and $E_1$(LO) modes are usually very weak due to the destructive interference between the deformation and the Fröhlich potentials [53]. We notice that the intensities of the AB, AMs, and LO mode increase with the increase of the Mn and Co nominal concentrations, evidencing its close relationship with the insertion of dopant in the ZnO structure. It can be observed that the intensities of the AB, AMs, and LO mode are considerably more significant for Mn-doped than that for Co-doped samples. Lower Mn concentrations lead to intensities as higher as those produced by larger Co concentrations.

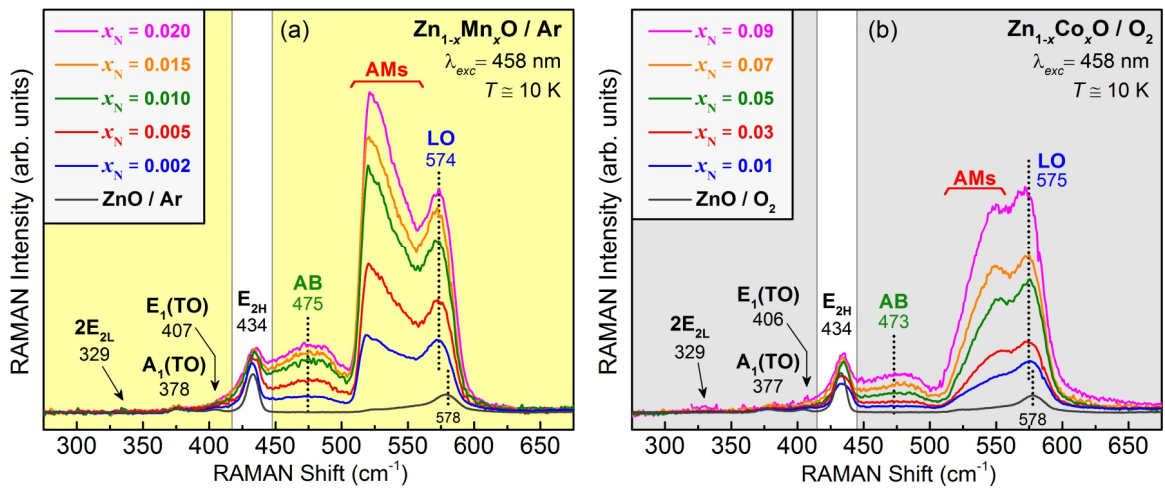

**Figure 3:** Raman scattering spectra of the (a) $Zn_{1-x}Mn_xO$ prepared in Ar and (b) $Zn_{1-x}Co_xO$ samples prepared in $O_2$ atmosphere. The spectra were acquired at 10 K and are normalized by the integrated area of the vibrational mode $E_{2H}$. The labeled ZnO sample corresponds to the reference undoped ZnO sample. The excitation was performed using the 458 nm laser line of an Ar/Kr laser. In (a) and (b) the vertical scales are the same for comparison.

For accurately identifying the observed vibrational modes and quantitatively determining the influence of the dopants on the ZnO structure, a Gaussian fit of the Raman spectra was performed for the two sets of samples. Figures 4(a) and 4(b) present the best-obtained fit for the samples with higher Mn ($x_N = 0.02$) and Co ($x_N = 0.09$) nominal concentrations, respectively. First, we consider the good agreement between the experimental data (symbols) and the Gaussian fit (continuous lines). For both samples, the additional band (AB) is centered at around 470 cm⁻¹, while the additional modes (AMs), located at the low-frequency side of the LO mode (500-550 cm⁻¹), were deconvoluted into two additional modes; one at ~526 cm⁻¹ (denoted as AM1 in Figure 2), and another at ~545 cm⁻¹ (AM2). It is also important to notice that the intensity of the AMs modes concerning the LO mode is higher for the Mn-doped samples than for the Co-doped samples (the left-shoulder of the broadband along 500-600 cm⁻¹ is higher for the Mn-doped samples, while the right-shoulder is higher for the Co-doped samples). For the $Zn_{1-x}Mn_xO$ set of samples, we could identify also a mode at ~517 cm⁻¹, it is labeled as $AM_{Mn}$ in Figure 4(a). This mode cannot be associated with the ZnO vibrational spectrum [16, 21, 52]. An analysis considering the nature of the $AM_{Mn}$ mode will be further presented.

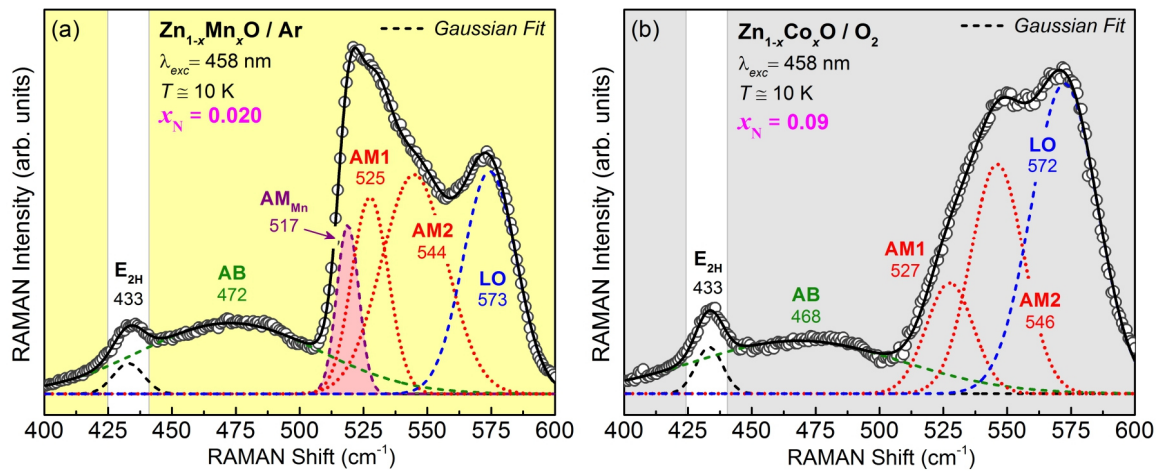

**Figure 4:** Raman spectra of the (a) $Zn_{1-x}Mn_xO$ ($x_N = 0.02$), and (b) of the $Zn_{1-x}Co_xO$ ($x_N = 0.09$). The symbols are the experimental points, the dashed/dotted lines are the Gaussian functions, and the solid lines correspond to the resulting fit. The spectra were acquired at 10 K and are normalized by the integrated area of the $E_{2H}$. The excitation was performed using the 458 nm laser line. In (a) and (b) the vertical scales are the same for comparison.

Figure 5 shows the evolution of the relative intensities ($I_R$) of the identified vibrational modes (Figure 4) concerning the $E_{2H}$ intensity as a function of the dopant nominal concentrations ($x_N$). The intensity corresponds to the integrated area obtained for each vibrational mode after the Gaussian fit of the spectra. We observe here a linear behavior for all modes. We also call attention to the linear coefficient values, all close to zero. By now, we can say that the incorporation of the dopants into the ZnO structure may be directly changing the electron and/or phonon density of states or, indirectly, through the introduction of structural disorders, making the one phonon density of states relatively more significant, leading to the observed linear increase in the intensity of the modes.

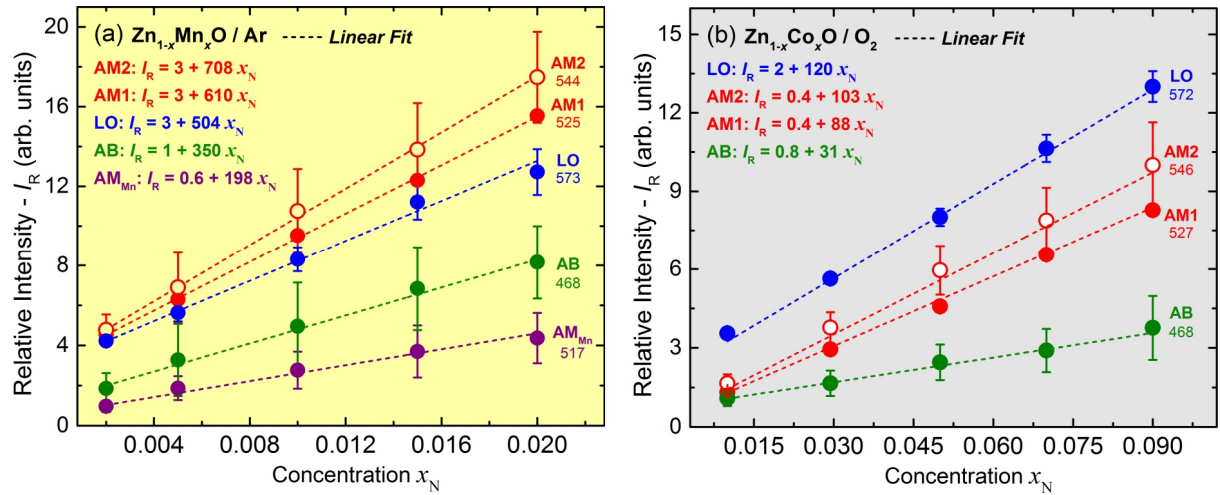

**Figure 5:** Relative intensities (integrated area) of the identified modes obtained after a Gaussian fit in the spectral range of 450 to 600 cm$^{-1}$ as a function of the nominal dopant concentration ($x_N$). For the (a) Zn$_{1-x}$Mn$_x$O and the (b) Zn$_{1-x}$Co$_x$O set of samples.

The emergence of additional modes in the range of 500-550 cm$^{-1}$ and the enhancement of the LO mode are always observed for implanted ZnO crystals with different elements [36, 37, 39, 54, 55]. Reuss *et al*. reported the broadband along 500-600 cm$^{-1}$ in Ga$^+$- and N$^+$-implanted ZnO single crystals and thin films [37]. After annealing the samples at a relatively high temperature (800 °C) the broadband was removed. This led them to associate the broadband with structural crystal damages promoted by the ion bombardment during the implantation process. In turn, Friedrich *et al*. studied the vibrational properties of H$^+$- and N$^+$-implanted ZnO single crystals at different excitation energy (458 and 633 nm). They observed the emergency of the broadband along 500-600 cm⁻¹ after the implantation and a resonant behavior increasing the excitation energy. Once the broadband was observed under sub-bandgap excitation, they attributed the nature of the broadband to an extrinsic Fröhlich interaction [36]. In this model, bound excitons are created due to impurities, which provide intermediate electronic states in the bandgap for resonant Raman scattering [56]. Besides, Ye *et al*. also linked the additional modes around 500-550 cm⁻¹ and the enhancement of the LO mode in the Raman spectra for As⁺-implanted ZnO single crystals to resonant Fröhlich optical phonons [54]. Sechogela *et al*. observed the same broadband in H⁺-implanted ZnO; they deconvoluted the broadband into the $A_1$(LO) mode at 579 cm⁻¹ and two modes, one at 525 cm⁻¹ (lower intensity) and the other at 553 cm⁻¹ (higher intensity) [55]. There are also reports showing the activation of these modes in undoped ZnO samples submitted to mechanical milling [44, 45]. Šćepanović *et al*., studying the effect of mechanical milling on undoped ZnO, also reported the presence of a broadband along 500-600 cm⁻¹ and its increase with the milling [44]. The authors deconvoluted the 500-600 cm⁻¹ broadband into the LO mode at ~580 cm⁻¹, and into two surface optical phonons (SOP) of Fröhlich character, one at ~515 cm⁻¹ (lower intensity), and another at ~550 cm⁻¹ (higher intensity). They addressed the increase of the LO mode to the increase in the concentration of point defects (oxygen vacancies), and the emergence of the SOP modes due to the increased structural disorder promoted by the mechanical milling on the surface of the ZnO grains. In fact, within the polariton description model for finite-size systems the boundary conditions lead to solutions for such SOP in the polariton equations ranging between the TO and LO region, depending on the geometry of the system [57]. In this approach the frequency of the SOP ($\omega_{SO}$) is given by [57]:

$$\omega_{SO}^2 = \frac{\varepsilon_0 + \varepsilon_M}{\varepsilon_\infty + \varepsilon_M}\, \omega_{TO}^2\,, \tag{1}$$

where $\omega_{TO}$ corresponds to the frequency of the TO modes, $\varepsilon_0$ and $\varepsilon_\infty$ are the material dielectric constants at low and high frequencies, while $\varepsilon_M$ is the dielectric constant of the surrounding media.

Considering the calculated dielectric tensor for the ZnO system as presented in Refs. [58, 59], assuming $\varepsilon_M = 1$, and the measured frequencies for the $A_1$(TO) at ~377 cm$^{-1}$ and $E_1$(TO) at ~410 cm$^{-1}$ (Figure (3)), Equation (1) returns SOP at the frequencies of ~525 and ~541 cm$^{-1}$, respectively. Therefore, the aspect of the broadband along 500-600 cm$^{-1}$ observed in our samples, the similarities of the AM1 and AM2 characteristics (intensity relation and frequencies) to those already reported and calculated, lead us to state that these modes are SOP of Fröhlich character induced by the doping and the increase of the surface area due to the existence of grains.

### 3.3 *Raman as a function of the excitation energy*

To check the Fröhlich character of the observed AB, AMs modes, and the enhancement of the LO intensity, we performed Raman scattering measurements with excitation at different wavelengths below the ZnO bandgap: 514 nm (2.41 eV), 488 nm (2.54 eV), and 458 nm (2.71 eV). Figure 6 presents representative spectra obtained for the undoped ZnO reference samples, for the $Zn_{1-x}Mn_xO$ sample with $x_N = 0.02$, and the $Zn_{1-x}Co_xO$ sample with $x_N = 0.09$. The spectra for the entire set of samples are presented in the supplementary file (Figures S3 and S4). The spectra were acquired at 10 K. The vertical scale is the same in all panels for comparison. For the undoped ZnO reference samples (Figures 6(a) and 6(b)) it is observed a relatively small increase in the intensity of the LO mode as the excitation wavelength/energy decreases/increases, indicating a resonant effect. For undoped ZnO, such an increase in intensity under resonance is usually observed for the LO mode [60]. It is noteworthy the emergency of a band at lower frequencies of the LO mode, in the same range for the AMs observed for the doped samples. For the doped samples, $Zn_{1-x}Mn_xO$ ($x_N = 0.02$) and $Zn_{1-x}Co_xO$ ($x_N = 0.09$), we observe a small increase in the AB band, and a significant increase in the broadband along 500-600 cm$^{-1}$ (AMs and LO) with the increase of the photon energy (Figures 6(c) and 6(d)), also evidencing a resonant effect. Compared to the spectra for the undoped ZnO reference samples, the resonant effect for the doped samples is considerably more significant. With increasing the dopant concentration, the resonant effect is also enhanced (Figures S3 and S4). Resonant Raman scattering is a well-known phenomenon that occurs when the excitation energy is of the order of, or greater than, the bandgap of the semiconductor material [61]. In this case, the virtual states that participate in the scattering process are replaced by the real ones, which increases the scattering cross section for phonons coupled with electrons via Fröhlich interaction, as is the case of the polar LO phonons [62]. Therefore, the observed resonance behavior confirms the Fröhlich character of the AMs. Moreover, the observed resonance under excitation below the bandgap of ZnO for the undoped and doped samples (Figure 6) is also indicative of the presence of states within the bandgap (extrinsic Fröhlich interaction) due to structural defects for undoped ZnO reference samples, and also, more significantly, due to the incorporation of Mn and Co into the ZnO lattice for the doped samples [13, 36, 56].

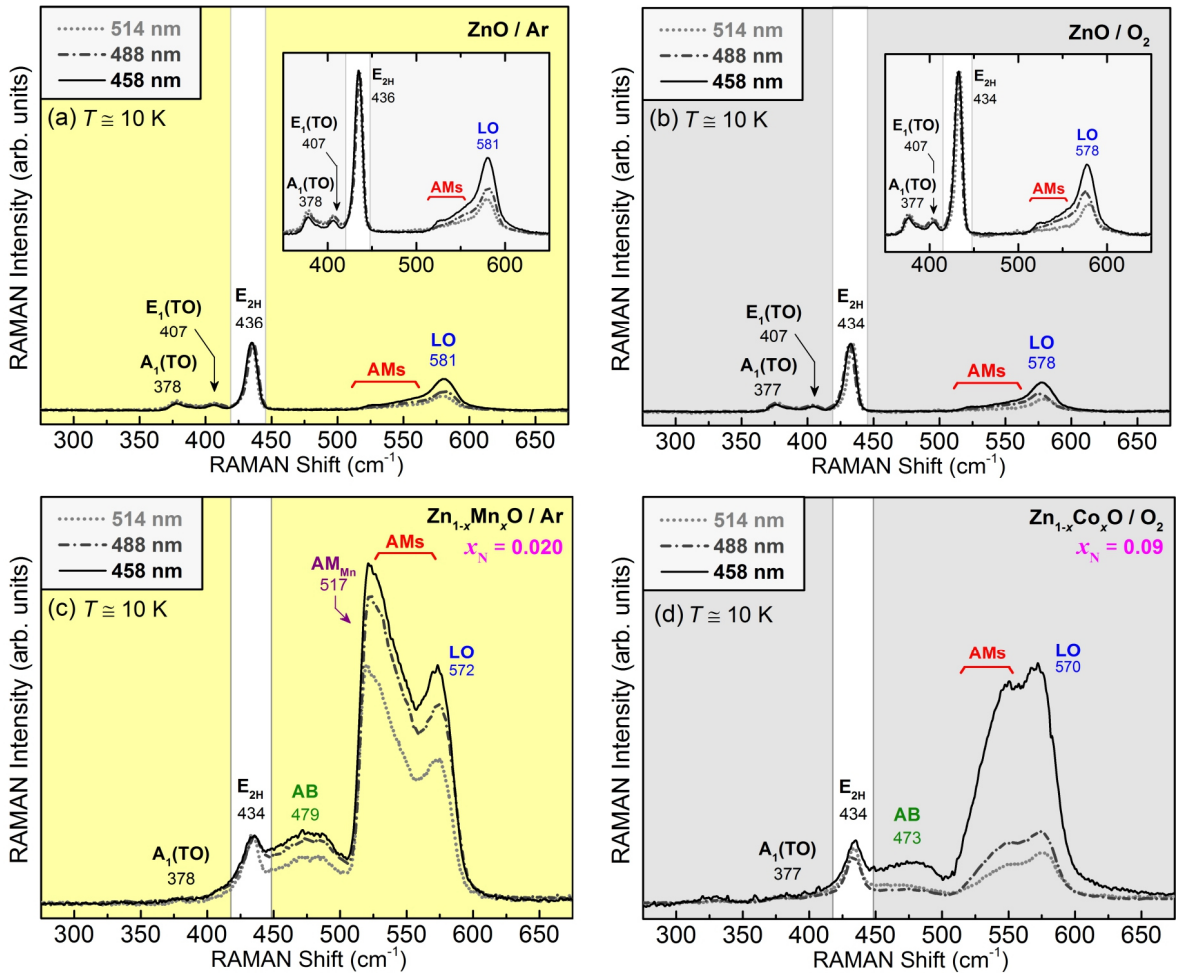

**Figure 6:** Raman spectra for the undoped ZnO reference samples prepared in (a) argon and in (b) oxygen atmospheres, (c) for the $Zn_{1-x}Mn_xO$ with Mn nominal concentration of $x_N = 0.020$; and (d) for the $Zn_{1-x}Co_xO$ with Co nominal concentration of $x_N = 0.09$. The excitation was performed using 514, 458, and 458 nm laser lines. The spectra are normalized by the integrated area of the $E_{2H}$. The vertical scale in all panels is the same for comparison.

Before proceeding, it is important to call attention to the fact that for the $Zn_{1-x}Mn_xO$ samples, the $AM_{Mn}$ mode at ~517 $cm^{-1}$ also resonates (Figure S5). Schumm *et al.*, after an extensive study on Mn implanted in ZnO bulk samples, reported a mode at this same frequency for Mn-doped ZnO samples and addressed it to a local vibrational mode (LVM) related to Mn inserted in the ZnO structure [39]. The same assignment was performed by other several authors [63, 64]. Besides, considering that the GaN and ZnO have the same structure, it is well known that Mn-doped GaN presents a Mn-related LVM at 667 $cm^{-1}$, between the $E_{2H}$ (569 $cm^{-1}$) and the $A_1(LO)$ (734 $cm^{-1}$) modes [65-67]. This mode was shown to have a linear-like dependence on the Mn concentration, and an intense polarization dependence similar to the one of the $A_1(LO)$ mode of GaN [65]. The observed $AM_{Mn}$ mode at ~517 $cm^{-1}$ in our Mn-doped samples is also located between the $E_{2H}$ and the LO modes (Figure 4(a)) and presents a linear correspondence to the Mn nominal concentration ($x_N$) (Figure 5(a)). However, despite the similarities between the characteristics of the Mn-LVM in GaN and the $AM_{Mn}$ in the $Zn_{1-x}Mn_xO$ samples, the observed resonance behavior excludes the possibility to address this mode as a Mn-LVM in the ZnO and reveals also its Fröhlich character.

Based on our data and in previous reports [39, 63, 64], it is clear that the observed vibrational mode at ~520 $cm^{-1}$ is related to the Mn incorporated into the ZnO host lattice. However, this mode cannot be attributed to Mn atoms incorporated in interstitial sites of the ZnO structure, since interstitial atoms are weakly linked and thus cannot vibrate at such high frequency [63]. Besides, considering a relatively high degree of Mn substitutional doping of the ZnO (2 at.%), it is plausible to infer it is possible to observe particular Mn-O vibrational modes in such samples. Julien *et al*. reported an extensive study on the vibrational properties of Mn oxides [68]. Recently, Bernardini *et al*. also reported a careful and comprehensive study on the Raman spectra of several different natural Mn oxides [69]. Among the Mn oxides, only MnO (manganosite) presents a relatively intense vibrational mode at ~529 $cm^{-1}$ [68, 69], close to the Mn-related mode observed in our samples. MnO has a cubic structure (space group *Fm3m*) with several Raman active modes where O vibrates against Mn. In particular, there are $A_1$ modes, where Mn-O vibrates as a one-dimensional chain (as the $A_1$ mode of ZnO). Thus, based on the experimental reported Raman spectra for MnO [68, 69], we would expect to observe the main Mn-O vibrational mode, as a normal mode in the ZnO lattice, at ~529 cm$^{-1}$. This normal mode, once having a LO character, resonates via Frohlich interaction, as observed in our Mn-doped samples (Figure S5). Therefore, in summary, the Mn incorporated into the ZnO lattice can be considered like as an isotopic disorder, but since they are coupled to each other via O atoms, the Mn-O vibrations can be considered a $A_1$(LO) normal mode of the ZnO lattice.

Some important and useful information can also be extracted from the spectra obtained under excitation above the ZnO bandgap. Figure 7 presents the spectra for the $Zn_{1-x}Mn_xO$ ($x_N = 0.020$), and $Zn_{1-x}Co_xO$ ($x_N = 0.09$) under excitation above the ZnO bandgap with a laser line of 351 nm (3.53 eV). A broad luminescence peak related to near band edge (NBE) electron transitions was subtracted from the spectra [13]. Under resonant Raman condition, the $A_1$(LO) phonon overtones at ~576 cm$^{-1}$ (upper inset in Figures 7(a) and 7(b)) dominate due to, as mentioned before, its polar character via intrinsic Fröhlich interaction with the excited electrons. The crystallinity and the order of the induced structural defects in our samples can be further evaluated by the electron-phonon interaction, which is very sensitive to the atomic scale disorder. The electron-phonon coupling constant ($\alpha$) can be probed under resonant excitation by the number of the measured overtones ($n$) [60]. We observe for the $Zn_{1-x}Mn_xO$ ($x_N = 0.02$) sample up to five overtones ($n = 5$), while only four overtones ($n = 4$) for the $Zn_{1-x}Co_xO$ ($x_N = 0.09$). These results indicate that the electron-phonon coupling strength is lower for the Co-doped ZnO sample, revealing for this sample a relatively lower crystallinity, corresponding to a higher structural disorder.

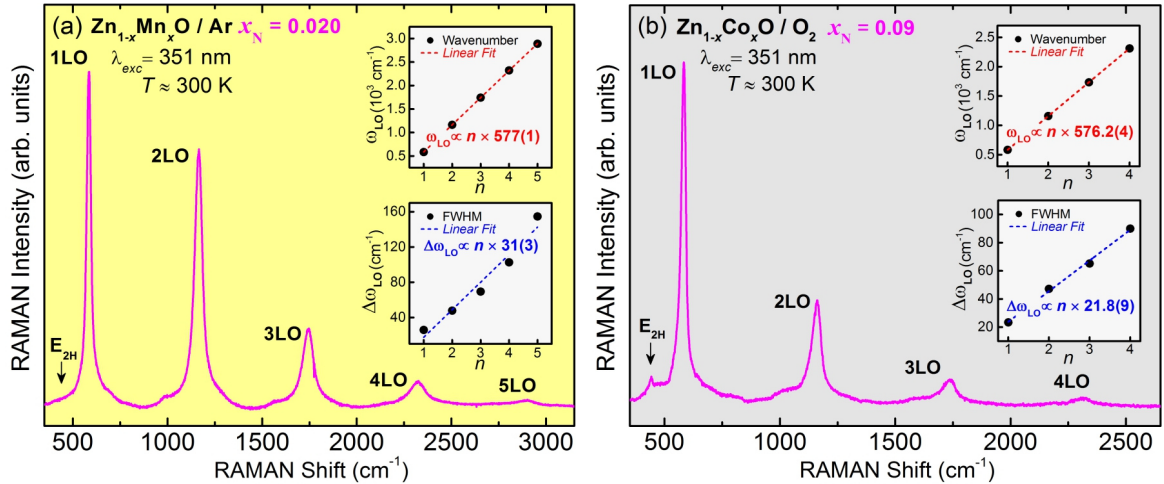

**Figure 7:** Resonant Raman scattering spectra for the (a) $Zn_{1-x}Mn_xO$ ($x_N = 0.020$) prepared in Ar and (b) $Zn_{1-x}Co_xO$ ($x_N = 0.09$) samples prepared in $O_2$. The spectra were acquired at room temperature (~300 K) and are normalized by the integrated area of the $E_{2H}$ mode. The excitation was performed using the 351 nm (3.53 eV, above the ZnO bandgap) laser line of an $Ar^+$ laser. The upper/lower inset presents the overtones central frequency/linewidths obtained after a Lorentzian fit.

In the ZnO structure, the Zn has a +2 oxidation state, coordination 4, and ionic radii of 0.74 Å. Considering substitutional doping of ZnO lattice (dopants taking the Zn tetrahedral site in the ZnO wurtzite structure), the $Mn^{2+}$ and $Co^{2+}$ crystal radii are 0.80 Å and 0.72 Å, respectively. Therefore, the $Mn^{2+}$ incorporation into the ZnO structure inevitably leads to significant structural distortions of the ZnO lattice [70]. On the other hand, the $Co^{2+}$ incorporation would not correspond to distortions of the same magnitude as those promoted by the $Mn^{2+}$ [21, 70]. Another important factor concerns the mass differences; the Co atomic mass (58.93) is closer to the Zn atomic mass (65.38) than that of Mn (54.94). Therefore, the $Co^{2+}$ insertion into the ZnO structure would affect, to a lesser extent, the vibrational properties of the system as compared to $Mn^{2+}$. After this analysis, the higher degree of structural disorder revealed in Figure 7 for the Co-doped sample can only be addressed to its higher dopant concentration (9 at.% for Co-doped ZnO *versus* 2 at.% for Mn-doped sample). Therefore, we cannot associate the intensity enhancement of the AB, AMs, and LO mode with such kind of structural disorder or, at least, it is less effective than the resonance effect presented before, once these vibrational modes are relatively much more intense for the Mn-doped samples (Figure 3), which show higher electron-phonon coupling constant corresponding to lower structural disorder (Figure 7).

### 3.4 *Bandgap engineering and structural defects analysis*

To confirm the hypothesis of the introduction of electronic states within the ZnO bandgap with Mn- and Co-doping, we performed diffuse reflectance measurements at room temperature in the range of 350 to 850 nm for the undoped ZnO reference samples and the doped samples $Zn_{1-x}Mn_xO$ ($x_N$ = 0.010 and 0.020), and $Zn_{1-x}Co_xO$ ($x_N$ = 0.05, and 0.09). Figures 8(a) and 8(b) present the obtained spectra, and Figures 8(c) and 8(d) present the correspondent Tauc plots [71]. A shift of the absorption edge is observed from the UV range of the electromagnetic spectrum for the undoped ZnO reference sample to the visible range for the doped ones. As the concentration of the dopant increases more pronounced the shift. The bandgap obtained for the samples via the Tauc plots reveals a significant reduction of the ZnO bandgap, from ~3.15 eV up to 2.37 eV for the $Zn_{1-x}Mn_xO$ ($x_N$ = 0.020) sample, and up to 2.62 eV for the $Zn_{1-x}Co_xO$ ($x_N$ = 0.09) sample. The absorptions in the visible region for the doped samples are attributed to multiple internal crystal-field transitions of the transition metal ions ($Mn^{2+}$ and $Co^{2+}$) replacing the $Zn^{2+}$ in the tetrahedral site of the ZnO wurtzite structure [72, 73]. For $Mn^{2+}$, the absorption corresponds to the overlap of *d-d* crystal-field transition of the high spin $Mn^{2+}$ $3d^5$ ion in tetrahedral oxygen coordination from its ground state $^6A_1(^6S)$ to $^4T_1(^4G)$, $^4T_2(^4G)$, $^4A_1(^4G)$, and $^4E(^4G)$ states (Figure 8(e)) [74, 75]. The $^6A_1(^6S)$ to $^4T_1(^4G)$ transition, the lowest energy transition, is of ~2.2 eV (2.37 eV for the sample $x_N$ = 0.02, Figure 8(c)), accounting for the characteristic reddish-orange color of the $Zn_{1-x}Mn_xO$ samples [74, 76]. In the case of $Zn_{1-x}Co_xO$ samples, it observed three well-defined absorption bands at 566 (2.19 eV), 612 (2.03 eV), and 659 nm (1.89 eV) also assigned to $Co^{2+}$ $3d^7$ *d-d* crystal-field transition in tetrahedral coordination from the ground state $^4A_2(^4F)$ to $^2A_1(^2G)$, $^4T_1(^4P)$ and $^2E(^2G)$ states, respectively (Figure 8(f)) [22, 77, 78]. The absorption edge for the Co-doped samples is ~2.62 eV for the sample $x_N$ = 0.09 (Figure 8(d)), corresponding to their characteristic dark-green color. The observation that the Mn and Co in the studied samples have +2 valence confirms the XPS results, and that they are located in tetrahedral sites also supports the previous structural analysis and leads us to conclude that the Mn and Co are substitutionally doping the wurtzite ZnO structure.

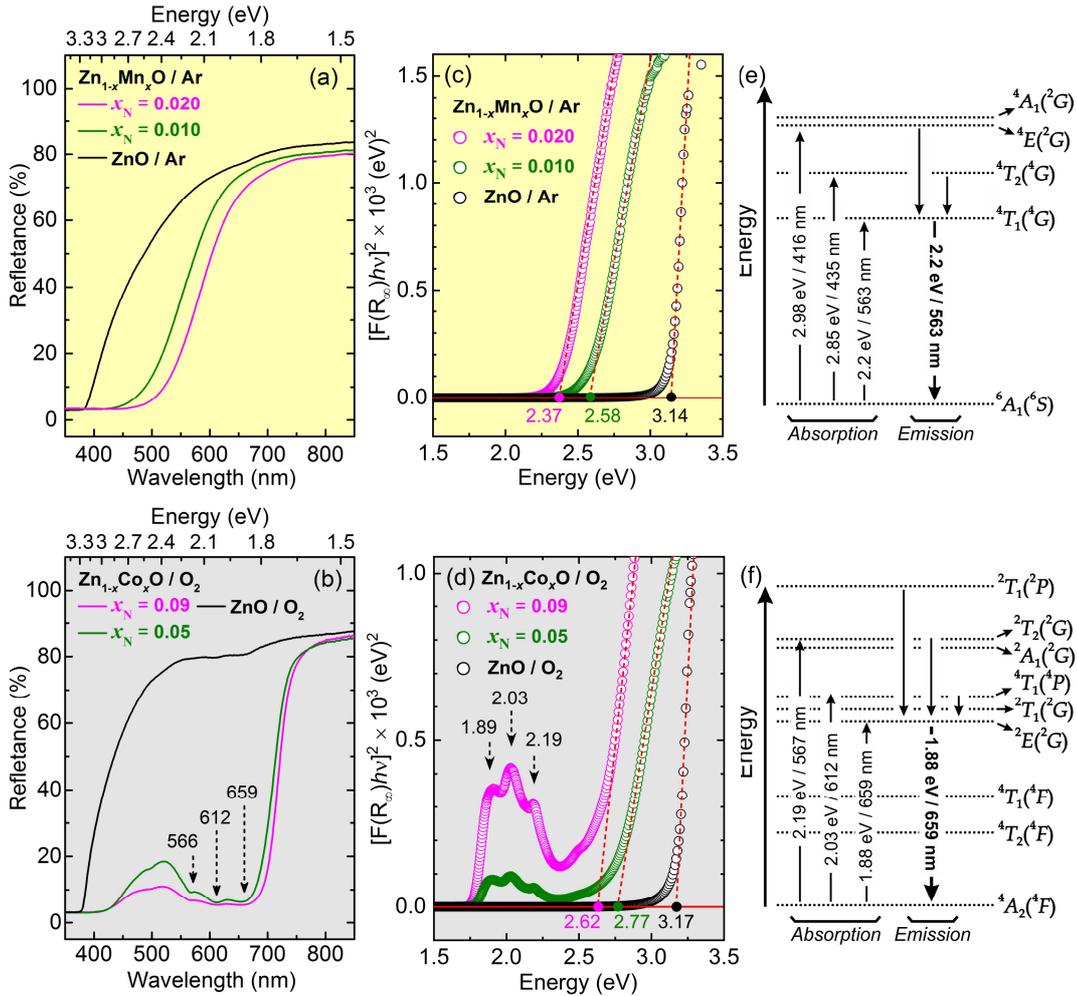

**Figure 8:** Reflectance spectra in the range from 350 to 850 nm for the (a) undoped ZnO and $Zn_{1-x}Mn_xO$ ($x_N = 0.010$, and 0.020) samples prepared in Ar, and for the (b) undoped ZnO and $Zn_{1-x}Co_xO$ ($x_N = 0.05$, and 0.09) samples prepared in $O_2$. (c) and (d) correspond to the Tauc plots of the reflectance spectra presented in (a) and (b), respectively. The dashed lines are linear fits of the absorption edges extrapolated to $h\nu = 0$ to find $E_g$. (e) and (f) present the energy diagram for the interatomic electron transition in the $Mn^{2+}$ and $Co^{2+}$ ions at a tetrahedral site (out off-scale), respectively.

These results confirm the insertion of dopant electronic states within the ZnO bandgap, which leads to an effective decrease of the bandgap and the intensity enhancement promotion of the Fröhlich modes due to a resonance effect (extrinsic Fröhlich interaction). Once the absorption edge shift is more pronounced for the Mn-doped samples, it can be the reason for the observed relatively higher intensities of the studied modes for this set of samples (Figure 3 and Figure 6). These results also lead us to point out that the observation of the broadband along 500-600 cm$^{-1}$ (AMs and LO) in the Raman spectra of the TM-doped ZnO are signatures of the effective substitutional doping of the structure and the tuning the ZnO bandgap from UV to lower wavelengths up to the visible part of the electromagnetic spectrum.

To gain more insight into this issue, the effects of the defects on the ZnO vibrational spectra, we also analyzed the vibrational spectrum of a mechanically milled $Zn_{1-x}Co_xO$ ($x_N = 0.05$) set of samples. The reduction of the sizes of the grains by fracture during the milling increases the surface/volume ratio, leading to an increase in surface effects. Mechanical milling is also a process that introduces structural defects, mainly on the surface of the powders. Figure 9 presents the obtained spectra for the milled $Zn_{1-x}Co_xO$ ($x_N = 0.05$) samples. Figure S6 shows the XRD patterns for the milled samples, only diffraction peaks related to the ZnO without undesired secondary phases are observed. Representative micrographs and the corresponding histograms for the entire set of samples are presented in Figures S7 to S11. We first observe that the milling does not affect the AB band. Besides, the intensity of the broadband along 500-600 cm$^{-1}$, with the enclosed AMs and LO mode, increases as the milling time increases. It is interesting to note that the relative intensity of the LO mode concerning the AMs modes also increases with the milling time, meaning that the LO mode is more affected by the milling process than the AMs. Once the AB is not affected by the mechanical milling, and remembering that the AB presents a linear dependence on the dopant concentration (Figure 5), we infer that the AB band is related to structural disorders promoted by the doping, i.e., lattice distortions due to the dopant incorporation. Besides, the enhancement of the AMs and LO mode with milling time indicates that these modes are related to surface effects [79, 80], including here surface structural point defects promoted by mechanical milling.

To evaluate the impact of the mechanical milling on the electronic structure of the samples diffuse reflectance measurements have been performed at room temperature (350-850 nm) and low temperature (5 K) PL under excitation of a He-Cd 325 nm laser line. Figures 9(b) and 9(c) present the obtained reflectance spectra for the milled samples and the corresponding Tauc plot, respectively. The reflectance spectra are pretty similar to the spectra for the $Zn_{1-x}Co_xO$ samples presented before (Figure 6(b)), indicating that mechanical milling does not affect the overall characteristics of the spectra. Figure 9(d) shows the measured PL spectra for the milled samples.

The spectra can be divided into three distinct emissions [77]: (i) an emission around 3.35 eV addressed to transition between states close to the ZnO conduction and valence band edges denominated as near band edge emission (NBE) (Figure 9(e)); (ii) a significant emission extending from approximately 3.20 eV to 1.9 eV associated with structural defect transitions; and (iii) the $Co^{2+}$ interatomic emission band at around 1.82 eV (Figure 9(f)). In the NBE region (Figure 9(e)), the band at 3.38 eV is associated with excitons bound to neutral donor defect states ($D°X$), while the band centered at 3.32 eV is associated with recombination of electrons bound to donor defects with free holes in the valence band (BF) [81]. In the $Co^{2+}$ region (Figure 9(f)), the emission band at 1.880 eV is associated with the already mentioned transition from $^2E(^2G)$ to the $^4A_2(^4F)$ ground state of $Co^{2+}$ in the $3d^7$ configuration in a tetrahedral crystalline field. The 1.880 eV emission is followed by its $E_{2L}$ (~14 meV) and $E_{2H}$ (~55 meV) phonon replicas.

Figure 9(d) shows that, with the mechanical milling time, the intensity of the defect broadband increases, revealing that the mechanical milling also introduces specific defect levels inside the ZnO bandgap that are not detected via diffuse reflectance. These defect levels, in the same way as the introduced dopant levels, increase the scattering cross-section of the Fröhlich phonons [62], in our case, the AMs and LO mode. Such kind of behavior was also reported for undoped [82], Mn-doped [12], and Co-doped [13] ZnO. We can say more, as the LO mode is more affected by the mechanical milling than the AMs modes, and that the enhancement of the LO mode is directly related to a resonance effect, not to surface effects, it is possible to infer that the insertion of the defect levels inside the ZnO bandgap (Figure 9(d)) contribute more to the observed enhancement of the broadband along 500-600 cm$^{-1}$ (AMs and LO) under mechanical milling (Figure 9(a)), than the decrease in the grain size (increase of the surface/volume ratio).

The previous results lead us to inquire about the grain size distribution of the $Zn_{1-x}Mn_xO$ (Figure 3(a)) and $Zn_{1-x}Co_xO$ (Figure 3(b)) sets of samples. Figures 10(a) and 10(b) present the obtained statistical results for the samples with higher dopant nominal concentration, $Zn_{1-x}Mn_xO$ ($x_N = 0.020$) and $Zn_{1-x}Co_xO$ ($x_N = 0.09$), respectively. The measured mean grain size ($d$) for the Zn$_{1-x}$Mn$_x$O ($x_N$ = 0.020) sample ($d$ = 8.1 µm) is almost half of that measured for the Zn$_{1-x}$Co$_x$O ($x_N$ = 0.09) sample ($d$ = 15.2 µm). Based on the assumption that the AMs are SOP of Fröhlich character, the observed higher intensities of these modes concerning the LO mode for the Zn$_{1-x}$Mn$_x$O set of samples, as compared to those for the Zn$_{1-x}$Co$_x$O set of samples (Figure 4), can be explained in terms of the higher grain surface area for the Zn$_{1-x}$Mn$_x$O samples.

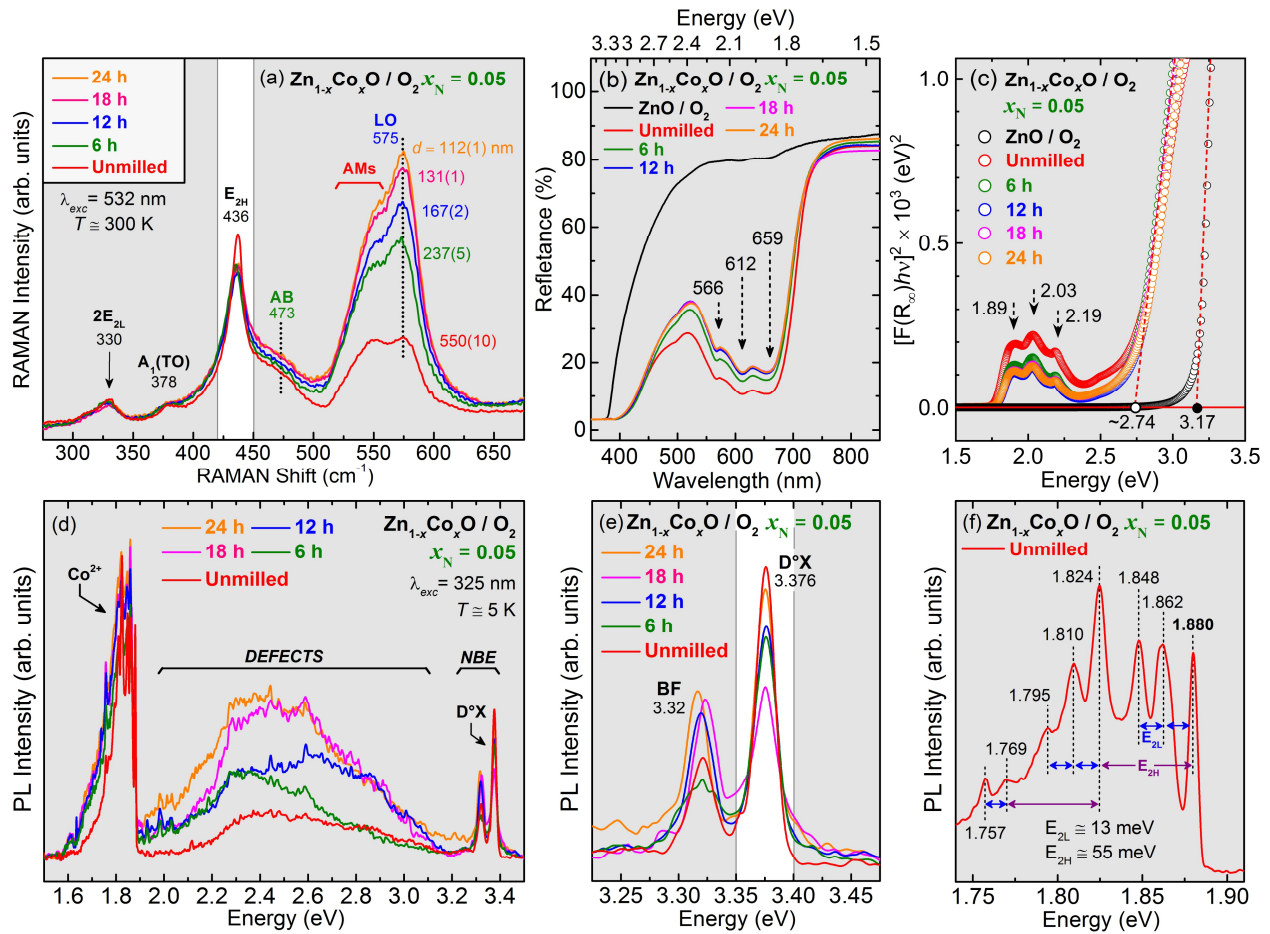

**Figure 9:** Spectra of the Zn$_{1-x}$Co$_x$O ($x_N$ = 0.05) samples prepared in O$_2$ and milled at different times. The main diameter ($d$ in nm) of the powder particles was evaluated via statistical analysis of the size distributions for each sample determined from SEM micrographs. The excitation was performed using a 532 nm solid-state laser line. The spectra were acquired at room temperature and are normalized by the integrated area of the E$_{2H}$. (b) Reflectance spectra of the Zn$_{1-x}$Co$_x$O ($x_N$ = 0.05) milled samples. (c) Correspondent Tauc plots of the reflectance spectra presented in (b). The dashed lines are linear fits of the absorption edges extrapolated to $h\nu$ = 0 to find $E_g$. (d) PL spectra obtained at 5 K for the milled Zn$_{1-x}$Co$_x$O ($x_N$ = 0.05) samples. The spectra are normalized by the integrated area of the denominated D°X emission band in the NBE region. Details of spectra in (e) the NBE region and in (f) the Co$^{2+}$ emission region for the unmilled reference sample.

Here it is pertinent to present some considerations about the differences in the ZnO grain growth under Mn and Co-doping. There are numerous studies on the ZnO grain growth kinetics;

these studies have revealed that the rate-controlling mechanism is the $Zn^{2+}$ diffusion, which proceeds via surface at the low-temperature regime [83] and via bulk at the high-temperature regime [78, 84, 85]. The metallic Mn-doping involves an oxidation process, considering that the Mn ions take the +2-oxidation state in the ZnO lattice [12, 17]. In Ar atmosphere, the metallic Mn incorporation occurs by the $Mn^{2+}$ sitting in the Zn site and forming oxygen vacancies ($V_O$). However, $V_O$ is a deep-donor defect, and its ionization is quite unlikely [13, 86]. In such a situation, the metallic Mn-doping of the ZnO lattice can be described by the defect chemical reaction presented in Equation (2), with no changes in the intrinsic ZnO $n$-type conductivity, but introducing $V_O$ point defects in the ZnO lattice. The $V_O$ defects lead to a Zn excess system, lowering the $Zn^{2+}$ mobility and decreasing the correspondent diffusion processes and grain growth rates.

Besides, the $Co_3O_4$ crystallizes in a spinel structure with $Co^{2+}$ and $Co^{3+}$ located at tetrahedral and octahedral sites, respectively [87, 88]. For a defect chemistry analysis, it can be considered a composition of CoO and $Co_2O_3$, $Co_3O_4 \leftrightarrow CoO + Co_2O_3$ [89, 90]. Therefore, the $Co_3O_4$ incorporation into the ZnO lattice in $O_2$ atmosphere can be described by the defect chemical reaction presented in Equations (3) and (4). Co-doping leads to a decrease of the ZnO $n$-type conductivity by introducing compensating holes ($h^•$) into the system and to the introduction of $V_{Zn}$ in the ZnO lattice. The reduced ZnO $n$-type conductivity under Co-doping via alloying ZnO with $Co_3O_4$ in the $O_2$ atmosphere is often reported [11, 13]. The presence of $V_{Zn}$ consequently favors the $Zn^{2+}$ diffusion process, promoting grain growth [78]. The mean diameter for the undoped ZnO sample sintered in $O_2$ is $d = 10.8$ μm (Figure S11), an intermediate value between those for the Mn-doped and Co-doped ZnO samples, supporting the previous assumptions.

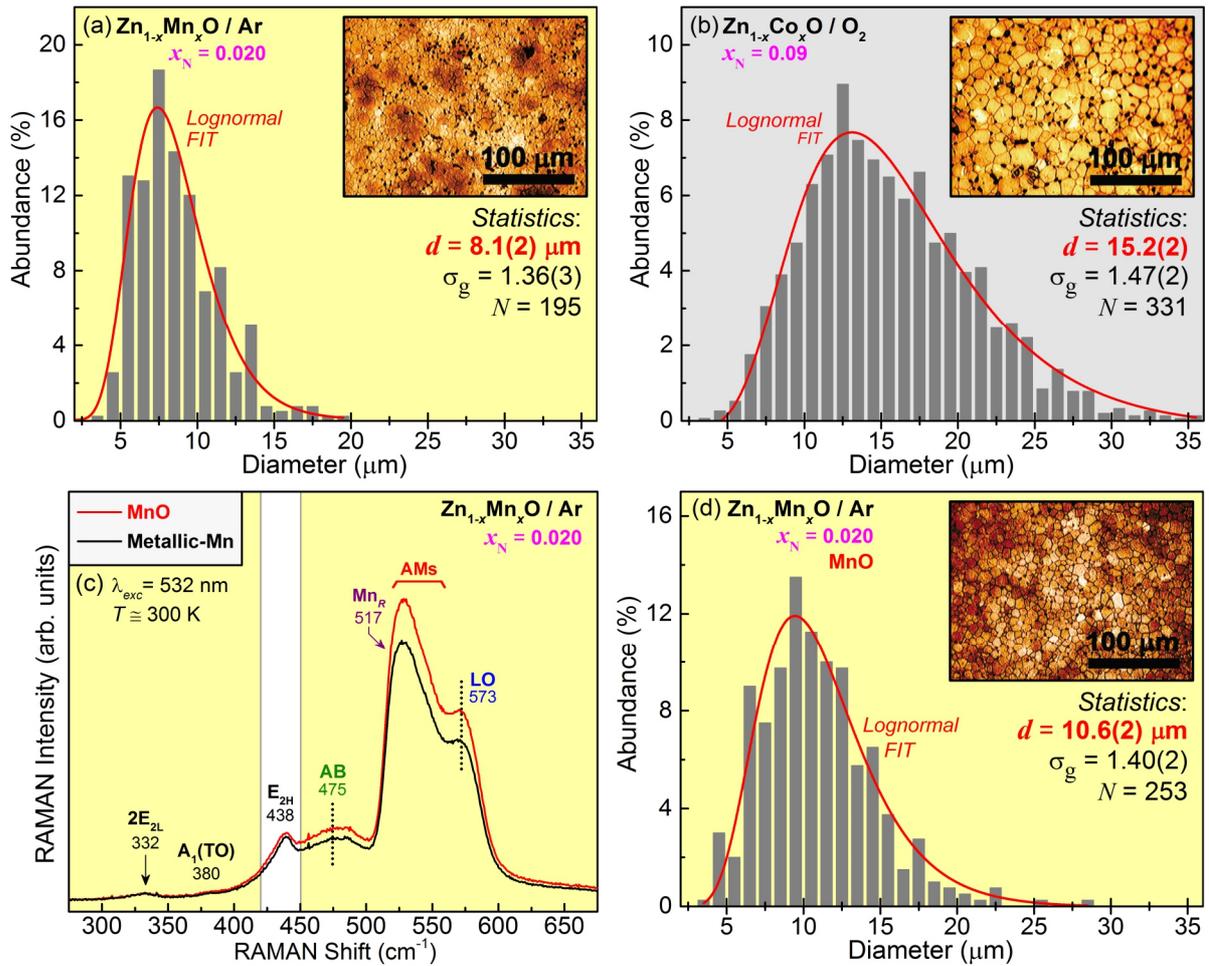

**Figure 10:** Particle size distribution histograms of the (a) Zn$_{1-x}$Mn$_x$O ($x_N$ = 0.020), and (b) Zn$_{1-x}$Co$_x$O ($x_N$ = 0.09) samples. The obtained statistical parameters after a lognormal fit of particle size distribution histograms for each sample are the mean grain diameter ($d$), the geometric standard deviation ($\sigma_g$), and the total number of counted grains ($N$). The inset presents a representative optical micrography of the polished surface of the samples. (c) Raman spectra for the Zn$_{1-x}$Mn$_x$O ($x_N$ = 0.02) samples prepared in Ar from different Mn precursors (MnO and metallic Mn). The excitation was performed using a 532 nm solid-state laser line. The spectra were acquired at room temperature and are normalized by the integrated area of the E$_{2H}$. (d) Particle size distribution histograms of the Zn$_{1-x}$Mn$_x$O ($x_N$ = 0.020) prepared in Ar from MnO precursor.

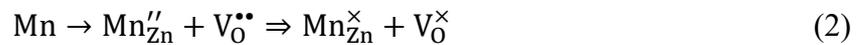

$$\text{Mn} \rightarrow \text{Mn}''_{\text{Zn}} + \text{V}_{\text{O}}^{\bullet\bullet} \Rightarrow \text{Mn}_{\text{Zn}}^{\times} + \text{V}_{\text{O}}^{\times} \tag{2}$$

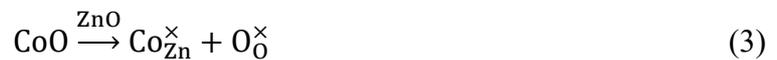

$$\text{CoO} \xrightarrow{\text{ZnO}} \text{Co}_{\text{Zn}}^{\times} + \text{O}_{\text{O}}^{\times} \tag{3}$$

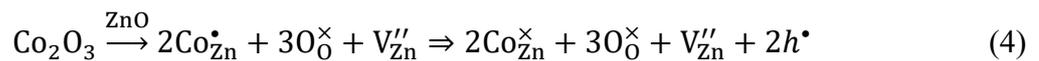

$$\text{Co}_2\text{O}_3 \xrightarrow{\text{ZnO}} 2\text{Co}_{\text{Zn}}^{\bullet} + 3\text{O}_{\text{O}}^{\times} + \text{V}''_{\text{Zn}} \Rightarrow 2\text{Co}_{\text{Zn}}^{\times} + 3\text{O}_{\text{O}}^{\times} + \text{V}''_{\text{Zn}} + 2h^{\bullet} \tag{4}$$

By using MnO as a precursor, Lage *et al.* recently showed that the incorporation of the Mn into the ZnO lattice takes place mainly at the surface of the ZnO grains with a solubility limit lower than 9 at.% [17]. In turn, at a similar synthesis condition, the Co solubility limit reaches almost 23 at. % [16]. After carefully analyzing the Mn incorporation dynamics into the ZnO, Santos *et al*. could increase the Mn solubility limit up to 21 at.% by using metallic Mn as a precursor and sintering the samples in Ar atmosphere [46]. To analyze the role of dopants on the surface vibrational properties of the doped ZnO, it was prepared, under the same conditions as the previously presented Mn-doped samples, another $Zn_{1-x}Mn_xO$ ($x_N = 0.020$) sample by using MnO (Alfa Aeser 99.99% purity) as the Mn precursor. Figure 10(c) compares the obtained Raman spectrum for this new sample with the spectrum for the $Zn_{1-x}Mn_xO$ ($x_N = 0.020$) sample prepared from metallic Mn. Figure 10(d) presents the grain size histograms for the sample prepared with MnO. We observe that the intensity of the AB is slightly higher for the sample prepared from MnO than those for the sample prepared from metallic Mn, while the AMs and LO mode are evidently larger for the MnO sample. Once that the MnO sample present a mean grain size diameter ($d = 10.6$ μm, Figure 10(d)) larger than that for the metallic Mn ($d = 8.1$ μm, Figure 8(a)), and that the MnO precursor leads to Mn-doped ZnO samples with relatively higher Mn concentration on the surface of the ZnO grains, we can also infer a relation of these modes (AB, AMs, and LO) with the elemental composition on the surfaces of the ZnO grains, which can be associated to structural distortions/disorders promoted by the doping on the surface of the ZnO grains. Based on this assumption, the observed higher intensities of the AB, AMs, and LO mode for the $Zn_{1-x}Mn_xO$ set of samples as compared to those for the $Zn_{1-x}Co_xO$ set of samples (Figure 4) can also be explained in terms of the difference between the $Mn^{2+}$ and $Co^{2+}$ solubilities into the ZnO structure. As the $Mn^{2+}$ presents a lower solubility, it incorporates more on the surface of the ZnO grains than the $Co^{2+}$, affecting largely the surface vibrational properties of the system, leading to higher intensities for the analyzed modes. Finally, Table 1 presents a summary of the observed vibrational modes for the studied set of samples, their symmetries, and the assignments, and a brief description of the corresponding process involved in each one.

**Table 1.** Room temperature frequencies and symmetries of the first- and second-order Raman spectra observed in the spectra of the $Zn_{1-x}Mn_xO$ and $Zn_{1-x}Co_xO$ set of samples and their assignments.

| *Frequency* (cm$^{-1}$) | | *Symmetry* | *Assignments* | *Process* |
|---|---|---|---|---|
| This work (average value) | From literature | | | |
| 329 | 330 [16] | $E_2$ | $2E_{2L}$ | Second-order |
| 377 | 378 [52] | $A_1$ | $A_1(TO)$ | First-order |
| 406 | 410 [52] | $E_1$ | $E_1(TO)$ | First-order |
| 434 | 438 [52] | $E_2$ | $E_{2H}$ | First-order |
| 470 | 475 [12], 465 [16] | - | AB | Dopant Related Surface Disorder |
| 517 | 519 [39] | $A_1$ | $AM_{Mn}$ | Mn Related ZnO Substutional doping |
| 526 | 525 [55] | $A_1$ | AM1 | Surface Optical |
| 545 | 553 [55] | $A_1$ | AM2 | Surface Optical |
| 572 | 574 [52] | $A_1$ | $A_1(LO)$ | First Order |

## 4  CONCLUSION

We presented a comprehensive analysis of the structure and optical properties of the $Zn_{1-x}Mn_xO$ and $Zn_{1-x}Co_xO$ samples (Mn- and Co-doped ZnO, respectively), focusing on the usual additional modes observed in the Raman spectral range of 400 to 700 cm$^{-1}$. The structural characterization confirms the Mn and Co are substitutionally doping the wurtzite ZnO lattice. With doping, besides the enhancement of the LO mode, the emergence of a broad band centered at ~470 cm$^{-1}$ (AB) and two additional modes at ~527 cm$^{-1}$ (AM1) and ~545 cm$^{-1}$ (AM2) were identified. For the Mn-doped samples, an extra additional mode located at ~517 cm$^{-1}$ ($AM_{Mn}$) has been found. The experimental results indicate that the AM1 and AM2 are surface optical phonons of Fröhlich character and that the AB band is related to structural disorders promoted by doping. A resonant behavior has been observed in $AM_{Mn}$ mode, but we have excluded to address it as a Mn-LVM, contrary to some reports in the literature. We argue that the $AM_{Mn}$ undoubtedly corresponds to Mn-O vibrations and that it could be considered as an A1(LO) normal mode of the ZnMnO lattice.

The promotion of the AB, AMs (SOPs), and LO mode and their intensity increase with the increase of the dopant (Mn and Co) concentration is explained in terms of three main effects. (i) The doping and the dopant accumulation on the surfaces of the ZnO grains promote the AB and AMs (SOPs) modes due to the introduction of structure disorder/distortions mainly on the surface of the ZnO grains. The intensity of these modes scales with the dopant concentration through the amount of caused structural disorder. (ii) The doping of the ZnO directly influences the grain growth kinetics, leading after sintering to different surface/volume ratios. Situations where grain growth is inhibited (final smaller grain sizes) lead to enhanced modes that are related to surface properties (AMs (SOPs)). (iii) The TM substitutional doping into the ZnO lattice corresponds to the introduction of the $3d$ levels inside the ZnO bandgap leading to the effective decrease of the ZnO bandgap and increasing the scattering cross section for phonons that are coupled with electrons via Fröhlich interaction and LO mode). In this context, inserting structural defects level inside de ZnO bandgap through ion bombardment or mechanical milling has a similar effect as the TM-doping.

Our results give valuable information on the vibrational properties of the TM-doped ZnO and the processing of these samples. It can be stated that the observation of the analyzed modes (activation of the AB and AMs (SOPs) and the enhancement of the LO mode) under doping is a signature of substitutional doping of the ZnO structure with the important and desired tuning of ZnO bandgap from the UV to the visible range of the electromagnetic spectrum, and with direct impact on the grain surface properties. Due to the similarities between the reported Raman spectra for other TM elements, besides the Mn and Co studied here, our conclusions can be generalized for the entire TM-doping elements of the ZnO structure.

## ACKNOWLEDGMENTS

Support from agencies CAPES is gratefully acknowledged (88881.064997/2014-01). This work was partially financed with projects Prometeo2020-016 and RTI2018-093711-B-I00, from the Generalitat Valenciana and the "Agencial Estatal de Investigación". The authors also acknowledge Prof. Dr. A. C. Doriguetto, coordinator of the Laboratório de Cristalografia of the Universidade Federal de Alfenas where the XRD measurements were performed.

**REFERENCES**

1.  Coll, M., et al., *Towards oxide electronics: A roadmap.* Applied Surface Science **482** (2019) 1-93.

2.  Zhang, T., et al., *Multi-component ZnO alloys: Bandgap engineering, hetero-structures, and optoelectronic devices.* Materials Science & Engineering R-Reports **147** (2022) 100661.

3.  Tuller, H.L. and S.R. Bishop, *Point defects in oxides: Tailoring materials through defect engineering.* Annual Review of Materials Research **41** (2011) 369-398.

4.  Chen, J., et al., *Pulsed laser deposition and characteristics of epitaxial non-polar m-plane $ZnO_{1-x}S_x$ alloy films.* Journal of Alloys and Compounds **773** (2019) 443-448.

5.  Smyth, D.M., *The effects of dopants on the properties of metal oxides.* Solid State Ionics **129** (2000) 5-12.

6.  von Wenckstern, H., et al., *Anionic and cationic substitution in ZnO.* Progress in Solid State Chemistry **37** (2009) 153-172.

7.  Tufte, O.N. and P.W. Chapman, *Electron mobility in semiconducting strontium titanate.* Physical Review **155** (1967) 796-802.

8.  Tietze, T., et al., *Interfacial dominated ferromagnetism in nanograined ZnO: a $\mu$SR and DFT study.* Scientific Reports **5** (2015) 8871.

9.  Xing, G.Z., et al., *Emergent ferromagnetism in $ZnO/Al_2O_3$ core-shell nanowires: Towards oxide spinterfaces.* Applied Physics Letters **103** (2013) 022402.

10. da Silva, R.T., et al., *Multifunctional nanostructured Co-doped ZnO: Co spatial distribution and correlated magnetic properties.* Physical Chemistry Chemical Physics **20** (2018) 20257-20269.

11. de Godoy, M.P.F., et al., *Evidence of defect-mediated magnetic coupling on hydrogenated Co-doped ZnO.* Journal of Alloys and Compounds **555** (2013) 315-319.

12. de Almeida, V.M., et al., *Room temperature ferromagnetism promoted by defects at zinc sites in Mn-doped ZnO.* Journal of Alloys and Compounds **655** ( 2016) 406-414.

13. de Godoy, M.P.F., et al., *Defect induced room temperature ferromagnetism in high quality Co-doped ZnO bulk samples.* Journal of Alloys and Compounds **859** (2021) 157772.

14. Bibes, M., J.E. Villegas, and A. Barthelemy, *Ultrathin oxide films and interfaces for electronics and spintronics.* Advances in Physics **60**( 2011) 5-84.

15. Keimer, B., et al., *From quantum matter to high-temperature superconductivity in copper oxides.* Nature **518** (2015) 179-186.

16. Mesquita, A., et al., *Dynamics of the incorporation of Co into the wurtzite ZnO matrix and its magnetic properties.* Journal of Alloys and Compounds **637** (2015) 407-417.

17. Lage, V.M.A., et al., *Influence of reducing heat treatment on the structural and magnetic properties of MnO:ZnO ceramics.* Journal of Alloys and Compounds **863** (2021) 158320.

18. Ozgur, U., et al., *A comprehensive review of ZnO materials and devices.* Journal of Applied Physics **98** (2005) 041301.

19.  Ozgur, U., D. Hofstetter, and H. Morkoc, *ZnO Devices and Applications: A Review of Current Status and Future Prospects.* Proceedings of the IEEE **98** (2010) 1255-1268.

20.  Shohany, B.G. and A.K. Zak, *Doped ZnO nanostructures with selected elements - Structural, morphology and optical properties: A review.* Ceramics International **46** (2020) 5507-5520.

21.  de Carvalho, H.B., et al., *Absence of ferromagnetic order in high quality bulk Co-doped ZnO samples.* Journal of Applied Physics **108** (2010) 033914.

22.  Mamani, N.C., et al., *On the nature of the room temperature ferromagnetism in nanoparticulate Co-doped ZnO thin films prepared by EB-PVD.* Journal of Alloys and Compounds **695** (2017) 2682-2688.

23.  Hariharan, C., *Photocatalytic degradation of organic contaminants in water by ZnO nanoparticles: Revisited.* Applied Catalysis A: General **304** (2006) 55-61.

24.  Akyol, A., H.C. Yatmaz, and M. Bayramoglu, *Photocatalytic decolorization of remazol red RR in aqueous ZnO suspensions.* Applied Catalysis B-Environmental **54** (2004) 19-24.

25.  Colon, G., et al., *Highly photoactive ZnO by amine capping-assisted hydrothermal treatment.* Applied Catalysis B-Environmental **83** (2008) 30-38.

26.  Lizama, C., et al., *Optimized photodegradation of Reactive Blue 19 on TiO2 and ZnO suspensions.* Catalysis Today **76** (2002) 235-246.

27.  Mijin, D., et al., *A study of the photocatalytic degradation of metamitron in ZnO water suspensions.* Desalination **249** (2009) 286-292.

28.  Samadi, M., et al., *Recent progress on doped ZnO nanostructures for visible-light photocatalysis.* Thin Solid Films **605** (2016) 2-19.

29.  Wibowo, A., et al., *ZnO nanostructured materials for emerging solar cell applications.* RSC Advances **10** (2020) 42838-42859.

30.  Zhao, J.X., et al., *Characterizing the role of iodine doping in improving photovoltaic performance of dye-sensitized hierarchically structured ZnO solar cells.* ChemPhysChem **14** (2013) 1977-1984.

31.  Zhao, J.X., et al., *Eu doping for hierarchical ZnO nanocrystalline aggregates based dye-sensitized solar cell.* Electrochemistry Communications **32** (2013) 14-17.

32.  Dong, J., et al., *Impressive enhancement in the cell performance of ZnO nanorod-based perovskite solar cells with Al-doped ZnO interfacial modification.* Chemical Communications **50** (2014) 13381-13384.

33.  Lanjewar, M. and J.V. Gohel, *Enhanced performance of Ag-doped ZnO and pure ZnO thin films DSSCs prepared by sol-gel spin coating.* Inorganic and Nano-Metal Chemistry **47** (2017) 1090-1096.

34.  Li, M.K., et al., *The S concentration dependence of lattice parameters and optical band gap of a-plane ZnOS grown epitaxially on r-plane sapphire.* Journal of Alloys and Compounds **630** (2015) 106-109.

35.  He, Y.B., et al., *Structural and optical properties of single-phase ZnO$_{1-x}$S$_x$ alloy films epitaxially grown by pulsed laser deposition.* Journal of Alloys and Compounds **587** (2014) 369-373.

36. Friedrich, F. and N.H. Nickel, *Resonant Raman scattering in hydrogen and nitrogen doped ZnO.* Applied Physics Letters **91** (2007) 111903.

37. Reuss, F., et al., *Optical investigations on the annealing behavior of gallium- and nitrogen-implanted ZnO.* Journal of Applied Physics **95** (2004) 3385-3390.

38. Chen, Z.Q., et al., *Production and recovery of defects in phosphorus-implanted ZnO.* Journal of Applied Physics **97** (2005) 013528.

39. Schumm, M., et al., *Structural impact of Mn implantation on ZnO.* New Journal of Physics **10** (2008) 043004.

40. Huang, G.J., et al., *Synthesis, structure, and room-temperature ferromagnetism of Ni-doped ZnO nanoparticles.* Journal of Materials Science **42** (2007) 6464-6468.

41. Hoang, L.H., et al., *Raman spectroscopy of Cu doping in $Zn_{1-x}Co_xO$ diluted magnetic semiconductor.* Journal of Raman Spectroscopy **40** (2009) 1535-1538.

42. Tringe, J.W., et al., *Enhanced Raman scattering and nonlinear conductivity in Ag-doped hollow ZnO microspheres.* Applied Physics A: Materials Science & Processing **109** (2012) 15-23.

43. Zuo, J., et al., *Sb-induced size effects in ZnO nanocrystallites.* Journal of Raman Spectroscopy **32** (2001) 979-981.

44. Scepanovic, M., et al., *Raman study of structural disorder in ZnO nanopowders.* Journal of Raman Spectroscopy **41** (2010) 914-921.

45. Ghose, S., et al., *Role of Zn-interstitial defect states on $d^0$ ferromagnetism of mechanically milled ZnO nanoparticles.* RSC Advances **5** (2015) 99766-99774.

46. Santos, F.V., *Studies on the Mn incorporation into the ZnO matrix: synthesis and structural characterization.* 2018, Univerisidade Federal de Alfenas: Alfenas-MG.

47. He, Y.B., et al., *Solubility limits and phase structures in epitaxial ZnOS alloy films grown by pulsed laser deposition.* Journal of Alloys and Compounds **534** (2012) 81-85.

48. Wagner, C.D., et al., *Handbook of X-ray photoelectron spectroscopy.* 1978 Minnesota: Perkin-Elmer Corporation.

49. Li, W., et al., *Enhanced visible light photocatalytic activity of ZnO nanowires doped with $Mn^{2+}$ and $Co^{2+}$ ions.* Nanomaterials **7** (2017) 20.

50. Calleja, J.M. and M. Cardona, *Resonant raman scattering in ZnO.* Physical Review B **16** (1977) 3753-3761.

51. Damen, T.C., S.P.S. Porto, and B. Tell, *Raman effect in zinc oxide.* Physical Review **142** (1966) 570-574.

52. Cusco, R., et al., *Temperature dependence of raman scattering in ZnO.* Physical Review B **75** (2007) 165202.

53. Callender, R.H., et al., *Dispersion of raman cross-section in CdS and ZnO over a wide energy range.* Physical Review B **7** (1973) 3788-3798.

54. Ye, J.D., et al., *Raman-active Frohlich optical phonon mode in arsenic implanted ZnO.* Applied Physics Letters **94** (2009) 011913.

55. Sechogela, T., et al., *2 MeV proton irradiation effects on ZnO single crystal.* Surface Review and Letters **21** (2014) 1450012.

56. Colwell, P.J. and M.V. Klein, *Wave vector dependence and numerical value of the scattering efficiency for the resonant Raman effect in CdS.* Solid State Communications **8** (1970) 2095-2100.

57. Cantarero, A., *Review on Raman scattering in semiconductor nanowires: I. theory.* Journal of Nanophotonics **7** (2013) 071598.

58. Petousis, I., et al., *High-throughput screening of inorganic compounds for the discovery of novel dielectric and optical materials.* Scientific Data **4** (2017) 160134.

59. Jain, A., et al., *Commentary: The Materials Project: A materials genome approach to accelerating materials innovation.* APL Materials **1** (2013) 011002.

60. Scott, J.F., *UV resonant Raman scattering in ZnO.* Physical Review B **2** (1970) 1209-1211.

61. Leite, R.C.C. and S.P.S. Porto, *Enhancement of Raman cross Section in CdS due to resonant absorption.* Physical Review Letters **17** (1966) 10-12.

62. Fröhlich, H., *Electrons in lattice fields.* Advances in Physics **3** (1954) 325-361.

63. Cerqueira, M.F., et al., *Raman study of insulating and conductive ZnO:(Al, Mn) thin films.* Physica Status Solidi (A): Applications and Materials Science **212** (2015) 2345-2354.

64. Gebicki, W., et al., *Raman scattering study of ZnO:Ti and ZnO:Mn bulk crystals.* Superlattices and Microstructures **38** (2005) 428-438.

65. Gebicki, W., P. Dominik, and S. Podsiadlo, *Lattice dynamics and Raman scattering from GaN:Mn crystals.* Physical Review B **77** (2008) 245213.

66. da Silva, J.H.D., D.M.G. Leite, and A.R. Zanatta, *Resonant excitation of Mn local vibrational modes in the higher order Raman spectra of nanocrystalline $Ga_{1-x}Mn_xN$ films.* Journal of Physics-Condensed Matter **20** (2008) 252201.

67. Devillers, T., et al., *Functional $Mn-Mg_k$ cation complexes in GaN featured by Raman spectroscopy.* Applied Physics Letters **103** (2013) 211909.

68. Julien, C.M., M. Massot, and C. Poinsignon, *Lattice vibrations of manganese oxides - Part 1. Periodic structures.* Spectrochimica Acta Part A: Molecular and Biomolecular Spectroscopy **60** (2004) 689-700.

69. Bernardini, S., et al., *Raman spectra of natural manganese oxides.* Journal of Raman Spectroscopy **50** (2019) 873-888.

70. Kolesnik, S., B. Dabrowski, and J. Mais, *Structural and magnetic properties of transition metal substituted ZnO.* Journal of Applied Physics **95** (2004) 2582-2586.

71. Tauc, J., *Optical properties and electronic structure of amorphous Ge and Si.* Materials Research Bulletin **3** (1968) 37-46.

72. Liu, C., F. Yun, and H. Morkoc, *Ferromagnetism of ZnO and GaN: A review.* Journal of Materials Science-Materials in Electronics **16** (2005) 555-597.

73. Kane, M.H., et al., *Magnetic properties of bulk $Zn_{1-x}Mn_xO$ and $Zn_{1-x}Co_xO$ single crystals.* Journal of Applied Physics **97** (2005) 023906.

74. Furdyna, J.K., *Diluted magnetic semiconductors.* Journal of Applied Physics **64** (1988) R29-R64.

75. Jin, Z.W., et al., *Blue and ultraviolet cathodoluminescence from Mn-doped epitaxial ZnO thin films.* Applied Physics Letters **83** (2003) 39-41.

76. Chikoidze, E., et al., *ZnO:Mn as a member of II-VI : Mn family.* Applied Physics A: Materials Science & Processing **88** (2007) 167-171.

77. Valerio, L.R., et al., *Preparation and structural-optical characterization of dip-coated nanostructured Co-doped ZnO dilute magnetic oxide thin films.* RSC Advances **7** (2017) 20611-20619.

78. da Silva, R.T., et al., *A comprehensive study on the processing of Co:ZnO nanostructured ceramics: Defect chemistry engineering and grain growth kinetics.* Journal of Materials Science & Technology **138** (2023) 221-232.

79. Ruppin, R., *Thermal fluctuations and raman-scattering in small spherical crystals.* Journal of Physics C: Solid State Physics **8** (1975) 1969-1978.

80. Ruppin, R., *Surface modes and infrared-absorption of coated spheres.* Surface Science **51** (1975) 140-148.

81. Loan, T.T., N.N. Long, and L.H. Ha, *Photoluminescence properties of Co-doped ZnO nanorods synthesized by hydrothermal method.* Journal of Physics D: Applied Physics **42** (2009) 065412.

82. Zeng, H.B., X. Ning, and X.M. Li, *An insight into defect relaxation in metastable ZnO reflected by a unique luminescence and Raman evolutions.* Physical Chemistry Chemical Physics **17** (2015) 19637-19642.

83. Whittemore, O.J. and J.A. Varela, *Initial sintering of ZnO.* Journal of the American Ceramic Society **64** (1981) C154-C155.

84. Senda, T. and R.C. Bradt, *Grain growth in sintered ZnO and ZnO-Bi$_2$O$_3$ ceramics.* Journal of the American Ceramic Society **73** (1990) 106-114.

85. Shin, S.-D., et al., *Effect of sintering atmosphere on the densification and abnormal grain growth of ZnO.* Journal of the American Ceramic Society **79** (1996) 565-567.

86. Janotti, A. and C.G. Van de Walle, *Native point defects in ZnO.* Physical Review B **76** (2007) 165202.

87. Jiang, J. and L.C. Li, *Synthesis of sphere-like Co$_3$O$_4$ nanocrystals via a simple polyol route.* Materials Letters **61** (2007) 4894-4896.

88. de Carvalho, H.B., et al., *Transport and magnetotransport transition of thin Co films grown on Si.* Physica Status Solidi A: Applied Research **201** (2004) 2361-2365.

89. Sabioni, A.C.S., et al., *First study of oxygen diffusion in a ZnO-based commercial varistor.* Defect and Diffusion Forum **289-292** (2009) 339-345.

90. Knobel, M., et al., *Magnetic and magnetotransport properties of Co thin films on Si.* Physica Status Solidi A: Applied Research **187** (2001) 177-188.

# SUPPLEMENTARY DATA

# On the vibrational properties of transition metal doped ZnO: surface, defect, and bandgap engineering

Viviane M. A. Lage[a,b], Carlos Rodríguez-Fernández[c,d], Felipe S. Vieira[b],

Rafael T. da Silva[b], Maria Inês B. Bernardi[e], Maurício M de Lima Jr.[f],

Andrés Cantarero[c] and Hugo B. de Carvalho[b*]

[a] *Universidade Federal de Ouro Preto – UFOP, 35400-000 Ouro Preto, MG, Brazil*

[b] *Instituto de Ciências Exatas, Universidade Federal de Alfenas – UNIFAL, 37130-000 Alfenas, Brazil.*

[c] *Institute of Molecular Science, University of Valencia, 22085, E-46071 Valencia, Spain.*

[d] *Faculty of Engineering and Natural Sciences, Tampere University, 33720 Tampere, Finland*

[e] *Instituto de Física de São Carlos, Universidade de São Paulo – USP, 13560-970 São Carlos, Brazil*

[f] *Materials Science Institute, University of Valencia, 22085, E-46071 Valencia, Spain.*

*Corresponding Author*:
* hugo.carvalho@unifal-mg.edu.br

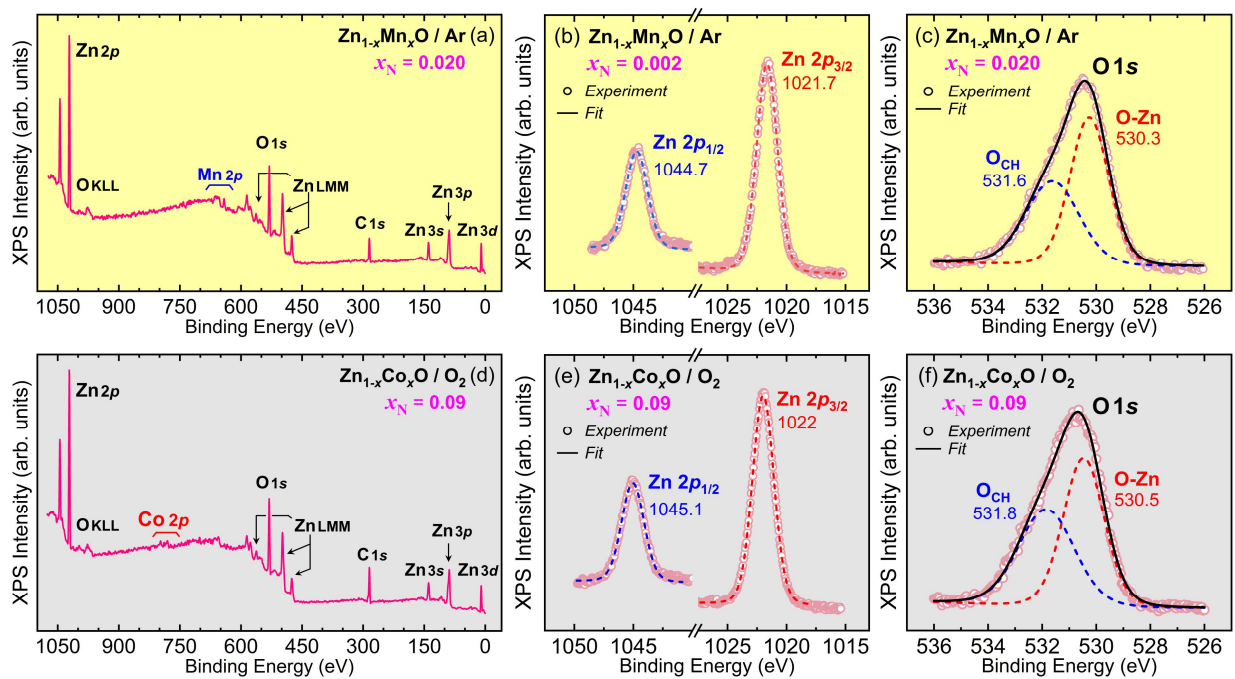

**Figure S1:** XPS spectra of the $Zn_{1-x}Mn_xO$ ($x_N = 0.02$) and the $Zn_{1-x}Co_xO$ ($x_N = 0.09$) samples: (a)/(d) survey, (b)/(e) $Zn\,2p$, and (c)/(f) O $1s$ core levels. The survey XPS spectra present peaks only at the binding energies corresponding to those of Zn, O, and Mn/Co core level energy states. No other impurities than carbon can be found in the spectra. The high-resolution peaks of the O $1s$ spectra for both samples can be divided into two peaks. The high intensity peak with a lower binding energy (~530 eV) is related to the $O^{2-}$ in the ZnO crystal lattice (O-Zn), while the broader peak at higher energy (~531.7 eV) is related to chemisorbed oxygens ($O_{CH}$) on the surface of the ZnO.

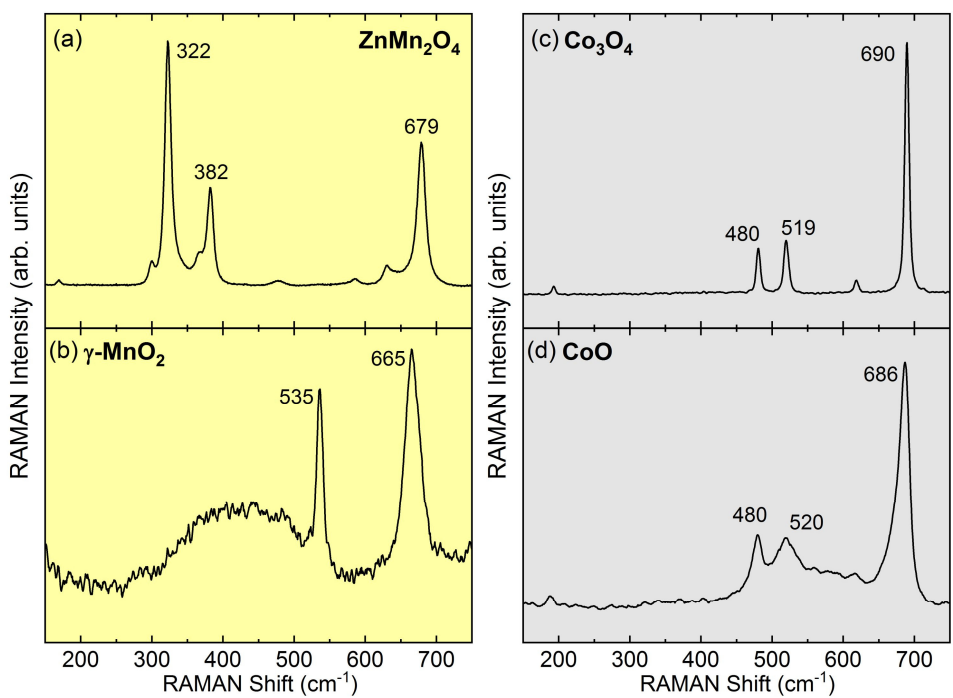

**Figure S2:** Raman spectra of the main secondary phases observed in the Zn-Mn-O and Zn-Co-O systems: (a) $ZnMn_2O_4$, (b) $\gamma$-$MnO_2$, (c) $Co_3O_4$, and (d) CoO. The spectra were acquired at room temperature by using a 532 nm laser as excitation.

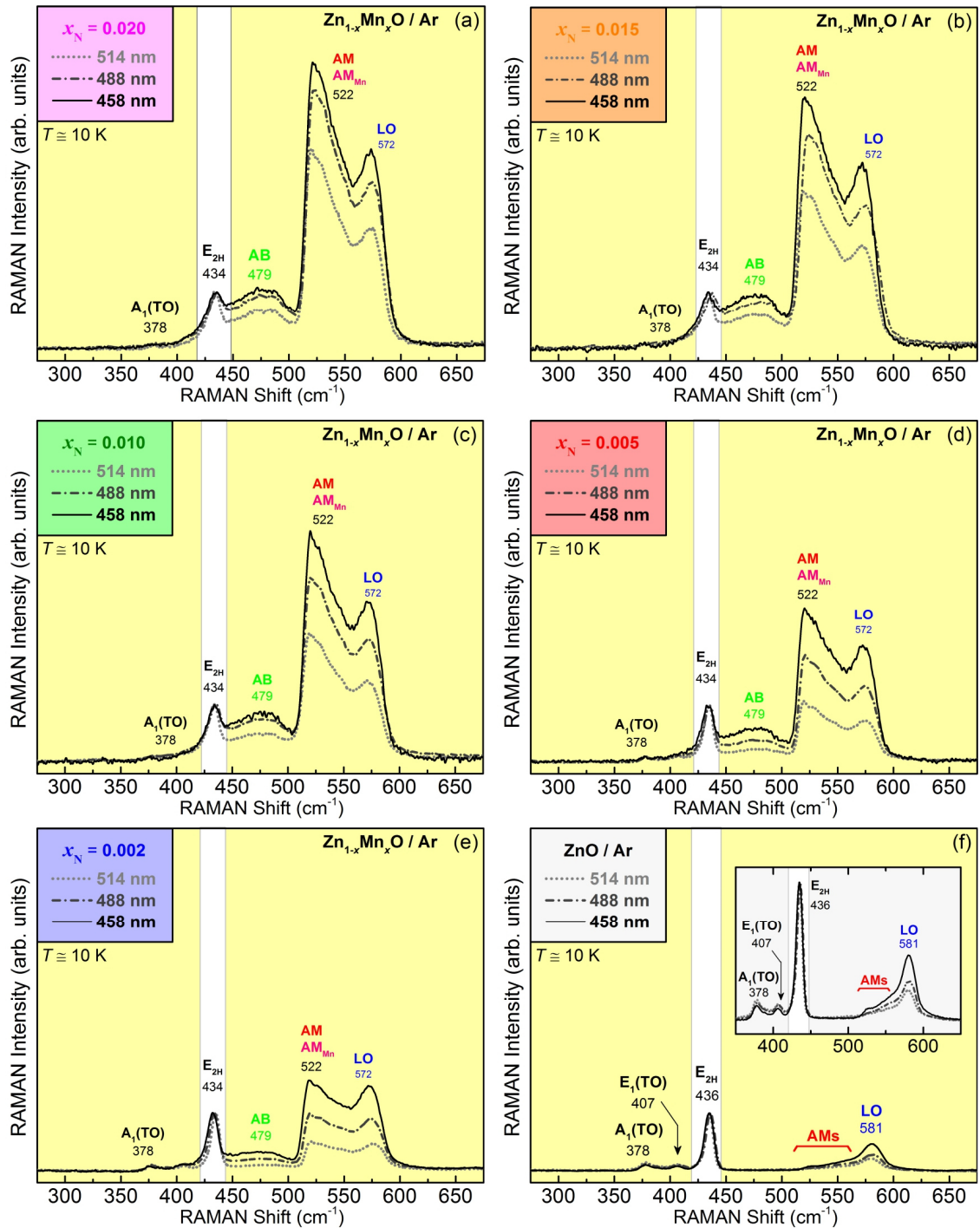

**Figure S3:** Raman spectra of de $Zn_{1-x}Mn_xO$ samples prepared in Ar at the nominal concentration of (a) $x_N = 0.020$; (b) 0.015; (c) 0.010; (d) 0.005; (e) 0.002 and (f) undoped ZnO. The spectra were acquired at the temperature of 10 K with excitation at the wavelengths of 514, 488, and 458 nm. The spectra are normalized by the integrated area of the $E_{2H}$ and have the same vertical scale for comparison.

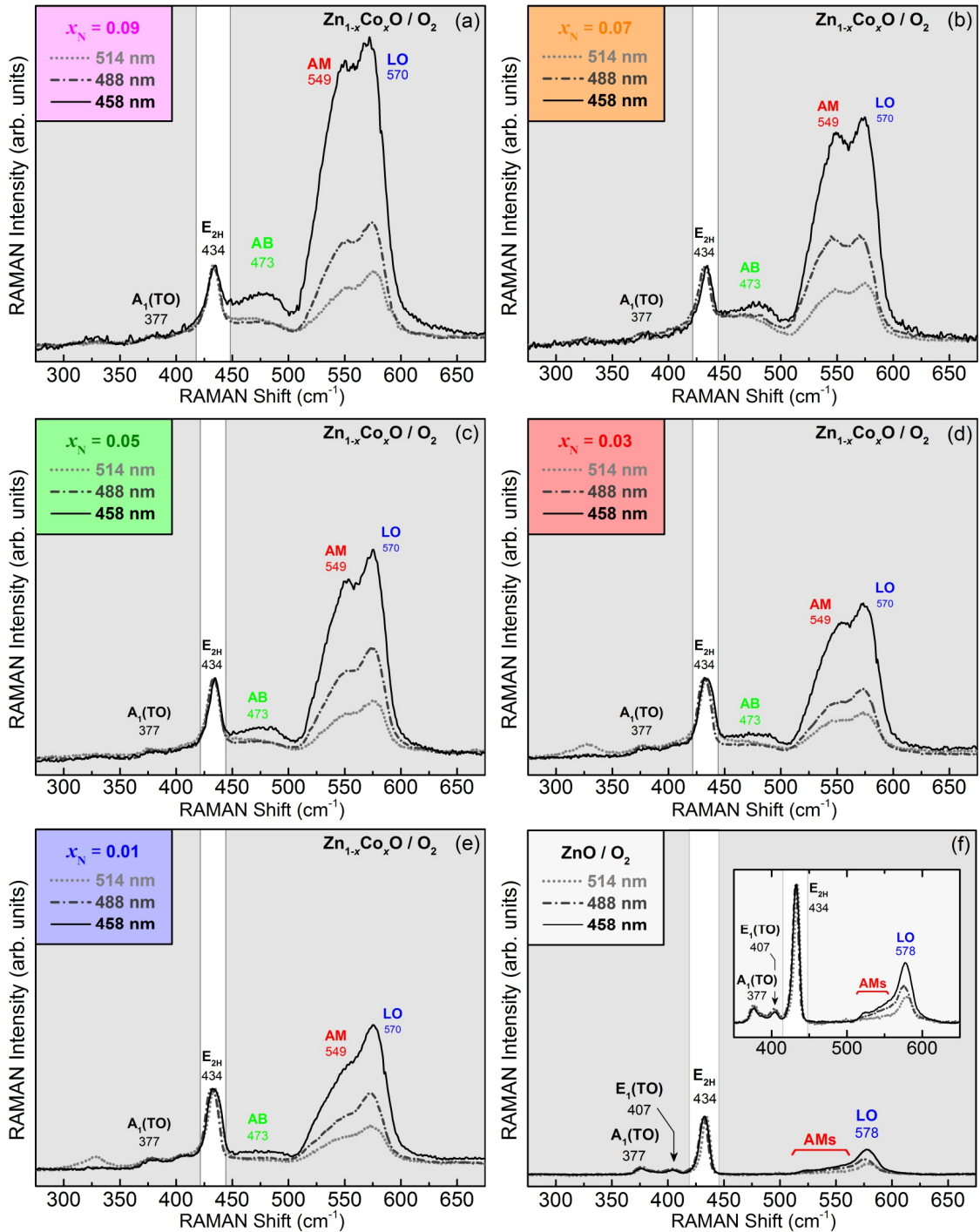

**Figure S4:** Raman spectra of de $Zn_{1-x}Co_xO$ samples prepared in $O_2$ at the nominal concentration of (a) $x_N = 0.09$; (b) 0.07; (c) 0.05; (d) 0.03; (e) 0.01 and (f) undoped ZnO. The spectra were acquired at the temperature of 10 K with excitation at the wavelengths of 514, 488, and 458 nm. The spectra are normalized by the integrated area of the $E_{2H}$ and have the same vertical scale for comparison.

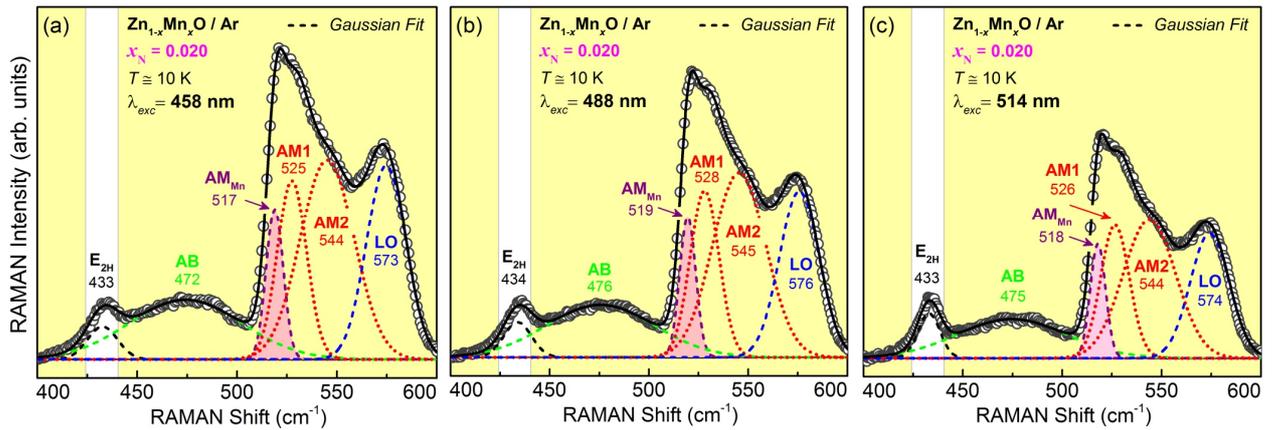

**Figure S5:** Raman spectra of de $Zn_{1-x}Mn_xO$ ($x_N = 0.020$) acquired at the temperature of 10 K with excitation at the wavelengths of (a) 458, (b) 488, and (c) 514 nm. The symbols are the experimental points, the dashed/dotted lines are the gaussian functions, and the solid lines correspond to the resulting fit. The spectra are normalized by the integrated area of the $E_{2H}$ and have the same vertical scale for comparison.

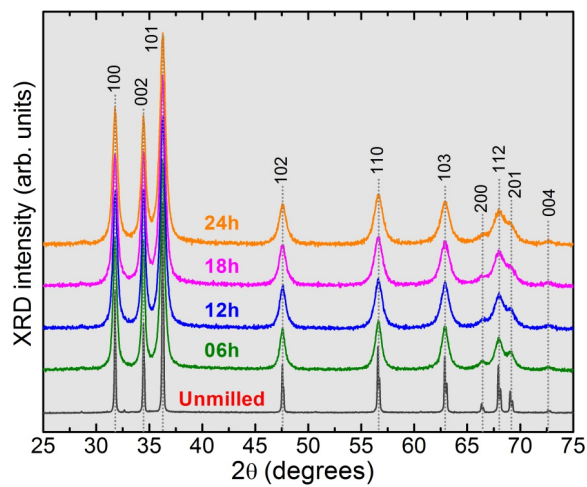

**Figure S6:** XRD pattern of the $Zn_{1-x}Co_xO$ ($x_N = 0.05$) samples sintered in $O_2$ and milled at different times.

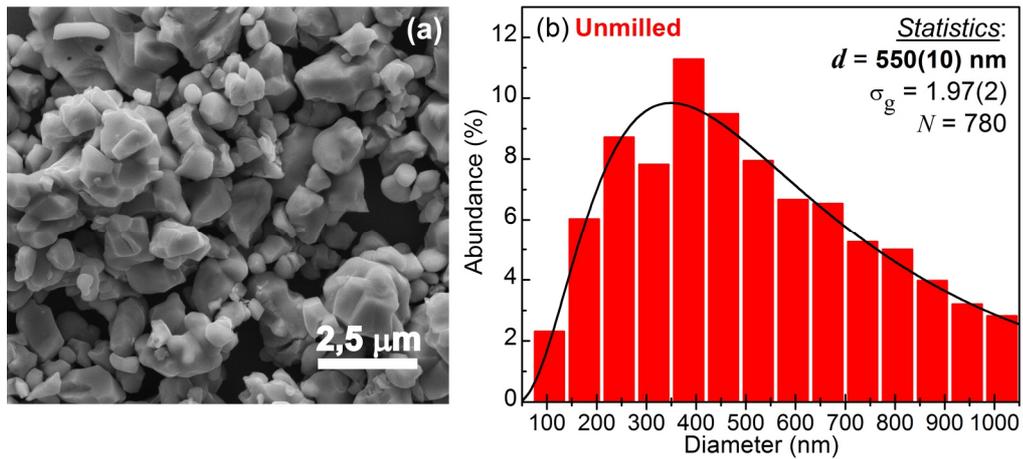

**Figure S7:** (a) Representative scanning electron micrographs of the unmilled $Zn_{1-x}Co_xO$ ($x_N = 0.05$) and (b) the correspondent grains size histogram distribution. A lognormal function fits the data.

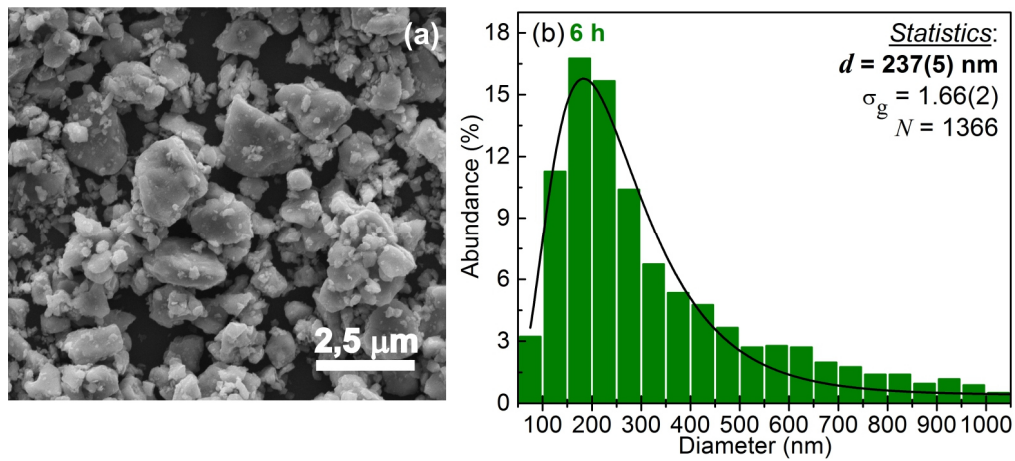

**Figure S8:** (a) Representative scanning electron micrographs of the $Zn_{1-x}Co_xO$ ($x_N = 0.05$) mechanically milled after 6 hours and (b) the correspondent grains size histogram distribution. A lognormal function fits the data.

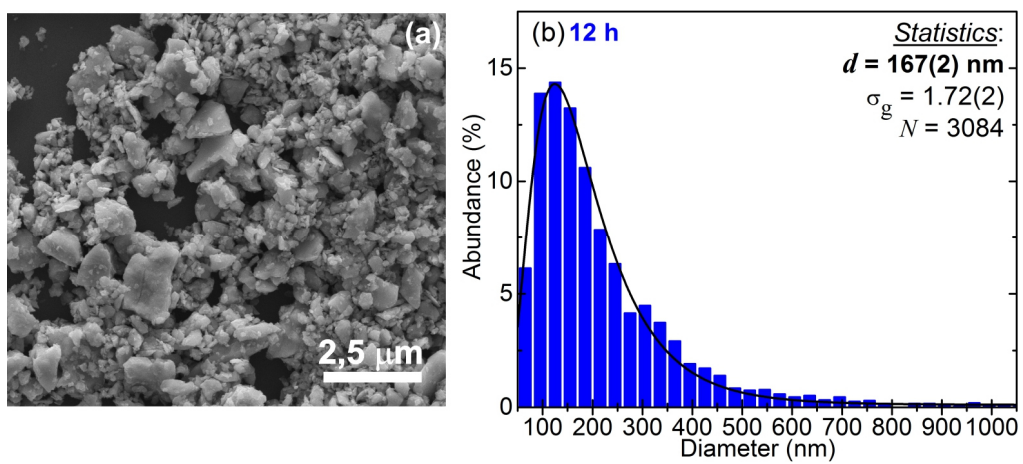

**Figure S9:** (a) Representative scanning electron micrographs of the $Zn_{1-x}Co_xO$ ($x_N = 0.05$) mechanically milled after 12 hours and (b) the correspondent grains size histogram distribution. A lognormal function fits the data.

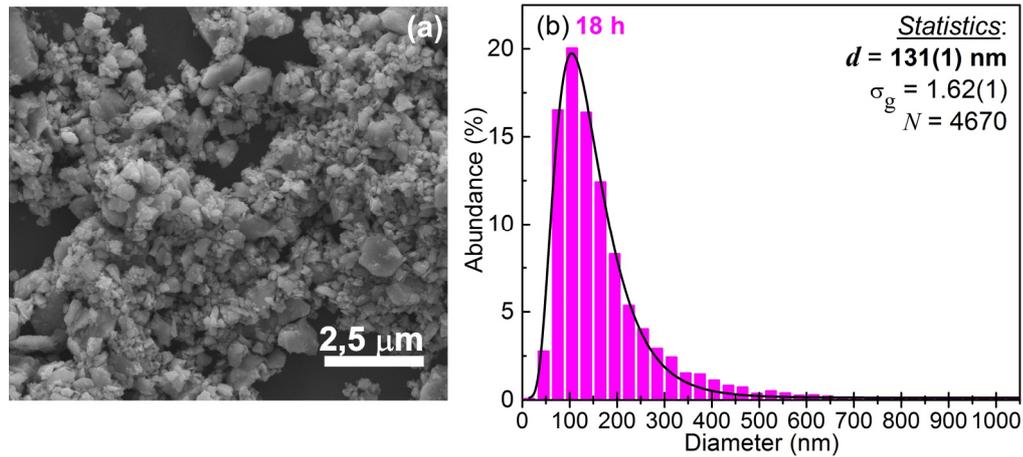

**Figure S10:** (a) Representative scanning electron micrographs of the $Zn_{1-x}Co_xO$ ($x_N = 0.05$) mechanically milled after 18 hours and (b) the correspondent grains size histogram distribution. A lognormal function fits the data.

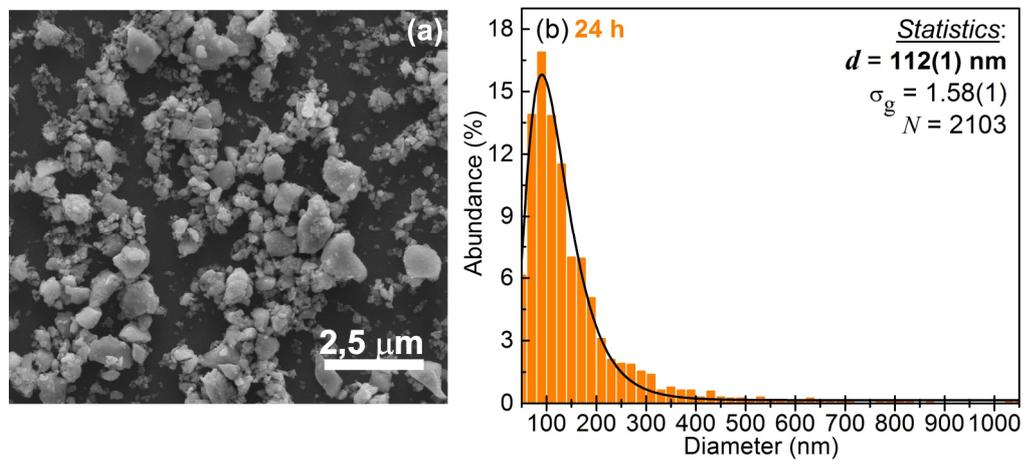

**Figure S11:** (a) Representative scanning electron micrographs of the $Zn_{1-x}Co_xO$ ($x_N = 0.05$) mechanically milled after 24 hours and (b) the correspondent grains size histogram distribution. A lognormal function fits the data.

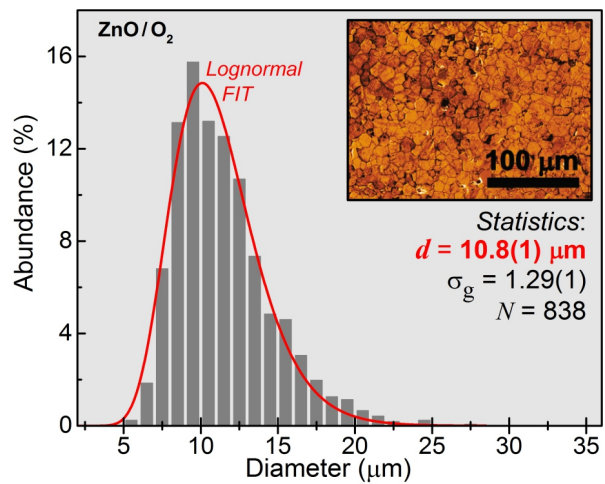

**Figure S12:** Particle size distribution histograms of the undoped $w$-ZnO sample prepared under $O_2$ atmosphere. The obtained statistical parameters after a lognormal fit of particle size distribution histograms for each sample are the mean grain diameter ($d$), the geometric standard deviation ($\sigma_g$), and the total number of counted grains ($N$). The inset presents a representative optical micrography of the sample polished surface.